\documentclass[prb,twocolumn,showpacs,preprintnumbers,amsmath,amssymb]{revtex4} 
\usepackage{graphicx}
\usepackage{dcolumn}
\usepackage{bm}
\usepackage{color}

\usepackage{float}
\usepackage{flafter}
\usepackage{epsfig,epstopdf}%

\allowdisplaybreaks

\newcommand{\rem}[1]{}
\begin{document}

\title{On the coarse-grained density and compressibility of non-ideal crystals: general theory and an application to cluster crystals}

\author{J. H\"aring$^\dagger$, C. Walz$^\dagger$, G.~Szamel$^\ddagger$ and M. Fuchs$^\dagger$}
\affiliation{$^\dagger$
Fachbereich f\"ur Physik, Universit\"at Konstanz, 78457 Konstanz, Germany\\
$^\ddagger$ Department of Chemistry, Colorado State University, Fort Collins, CO 80523, USA}

\date{\today}

\begin{abstract}
The isothermal compressibility of a general crystal is analyzed within classical density functional theory. Our approach can be used for homogeneous and unstrained crystals containing an arbitrarily high density of local defects. We start by coarse-graining the microscopic particle density and then obtain the long wavelength limits of the correlation functions of elasticity theory and the thermodynamic  derivatives.  We explicitly show that the long  wavelength limit of the microscopic density correlation function differs from the isothermal compressibility. It also cannot be obtained from the static structure factor  measured in a scattering experiment. We apply our theory to  crystals consisting of soft particles which can multiply occupy lattice sites ('cluster crystals'). The multiple occupancy results in a strong local disorder over an extended range of temperatures. We determine the cluster crystals' isothermal compressibility, the fluctuations of the lattice occupation numbers and their correlation functions, and the dispersion relations. We also discuss their low-temperature phase diagram.\end{abstract}



\maketitle

\section{Introduction}

In crystals, where translational invariance is spontaneously broken, strain enters as additional thermodynamic variable in the free energy, describing the distortion of the solid. The trace of the strain tensor is connected to the change in density. In particular, in ideal crystals, where all atoms can be unambiguously assigned to lattice sites and all lattice sites are occupied, density change is determined by the trace of the strain tensor. In real crystals, point defects like interstitials and vacancies are present, and density can change by both deformation of the solid (captured by the strain), and diffusion of defects \cite{Chaikin95}. Thus, the presence of defects opens the question how density and strain fluctuations are defined in real crystals. Here, no one-to-one mapping of atoms to lattice positions is possible. Therefore, the displacement field, whose symmetrized (in linear approximation) gradient gives the strain, cannot be obtained from the displacements of individual atoms from their lattice positions. Only recently microscopic definitions of strain and density fluctuations in real crystals were derived from the statistical mechanical description of real crystals, overcoming this difficulty \cite{Walz10}. This  work followed an earlier suggestion by Szamel and Ernst \cite{Szamel93,Szamel97}.  Preliminary Monte Carlo simulations and comparisons with older approaches, including to amorphous solids, indicated the potential of the microscopic theory \cite{Fuchs2012}.

An intriguing finding of the microscopic approach of  Ref.~[\onlinecite{Walz10}] concerns the coarse-grained density field $\delta n(\mathbf{r},t)$ which enters into the theory of crystal elasticity \cite{Martin72}. 
Even for arbitrarily large wavelengths, particle density fluctuations with wavevectors close to all (finite) reciprocal lattice vectors contribute to the coarse-grained density field. 
In this contribution, we discuss this at first surprising finding within the framework of density functional theory. This theory allows us to properly link microscopic and macroscopic density fluctuations  in states with broken translational symmetry in order to parallel the coarse-graining of the free energy functional achieved previously for \textit{e.g.}~homogeneous liquid crystals \cite{Stecki}.

Based on the microscopic definition of the coarse-grained variables of elasticity theory, we can address another  intriguing question, originally raised by Stillinger \cite{Stillinger66,Goetze67,Goetze68}. Namely, whether the structure factor is an analytic function around zero wavevector and whether its small wavevector limit coincides with the compressibility? We find that due to the long-ranged displacement correlations,  the small wavevector limit of the correlation function of the coarse-grained density field is non-analytic and depends on the direction relative to the crystal lattice.  We derive these results from density functional theory and can thus put them on a firm microscopic basis. Thus, we generalize earlier findings obtained within the harmonic crystal approximation \cite{Kayser81}. Because of the non-analyticity, special care is required when discussing the thermodynamic limit. From studies on two-dimensional crystals it is known that defects enter the expression for the isothermal compressibility in a complicated fashion \cite{Zippelius80}. We generalize these results to crystals of arbitrary symmetry. Correcting the appendix of Ref.~[\onlinecite{Walz10}], we also derive relations between fluctuation functions and thermodynamic derivatives. These results suggest that the elastic constants of crystals with point defects \cite{Pronk04} could be measured by microscopy techniques applied to colloidal crystals \cite{Reinke07}.

In order to test the theory, we apply it to so-called `cluster crystals' \cite{Mladek06,Mladek07} which consist of particles interacting with a soft-core repulsion. The softness of the potential allows for multiple occupancy of individual lattice sites by the particles and for fluctuations of the lattice sites occupation numbers. These fluctuations play the role of mobile local defects. Indeed, the approximation which considers these cluster crystals as ideal crystals (with a uniquely occupied lattice sites) is valid only at extremely low temperatures \cite{Likos07}, and the different crystal structures can only be understood by allowing for a distribution of site occupation numbers \cite{Wilding13,Zhang12}. For these crystals, we will derive thermodynamic derivatives, correlation functions, and dispersion relations, which were not accessible previously, and we will discuss their low temperature phases.

The paper is organized as follows: in Sect.~II we first recall definitions and results from Ref.~[\onlinecite{Walz10}] and then, in Sect.~III, we derive expressions for the fluctuations of displacement and density fields in an unstressed reference state. They are given by microscopic quantities defined in terms of the direct correlation function of the crystal. To facilitate application of these expressions,  we also invert these relations considering two sets of independent fluctuations, coarse-grained density and displacement field or defect density \cite{Fleming76} and displacement field. In Sect.~IV we derive the thermodynamic free energy, including the thermodynamic elastic susceptibilities, by coarse-graining the microscopic classical density functional. As the first step, we obtain the free energy functional containing the elastic fields, which reduces to the thermodynamic one for homogeneous fields. This is followed by the discussion of thermodynamic derivatives.  In Sect.~V we discuss the small wavevector limit of the coarse-grained density fluctuation function and show that it differs from the isothermal compressibility $\kappa$. We also discuss scattering functions and conclude that scattering experiments do not allow to measure the compressibility in a crystal, in contrast to liquids and gases \cite{Berne00}. Finally, in Sect.~VI we apply our theory to cluster crystals. We show that a simple mean-field density functional leads to surprisingly accurate values of compressibilities and occupation number fluctuations. Details of some of the calculations are presented in appendices.

\section{Coarse-grained fields}\label{sectII}

Crystals exhibit spontaneously broken translational symmetry (\textit{e.g.} the average density is non-uniform) and this, \textit{via} the Goldstone theorem, leads to long-ranged correlations. Specifically, the vector displacement field ${\bf u}({\bf r},t)$ possesses correlations which decay like the inverse distance. 
In ideal crystals, one can use the familiar expression for the microscopic density of the displacement field $\sum_i \mathbf{u}_i(t) \delta (\mathbf{r}-\mathbf{R}_i)$, involving the displacement $\mathbf{u}_i(t) = \mathbf{r}_i(t) -\mathbf{R}_i$ of the instantaneous position of the particle $i$, $\mathbf{r}_i(t)$, from its lattice site $\mathbf{R}_i$. However, in real crystals, in which defects are present, this expression is invalid \cite{Szamel93}. In order to find the microscopic definition for the displacement field ${\bf u}({\bf r},t)$ and for the other fields of elasticity theory, an alternative approach was developed in Ref.~[\onlinecite{Walz10}].

Before we discuss the approach of Ref.~[\onlinecite{Walz10}], we need to define precisely various
fields used in the present paper. First, we have microscopic densities, \textit{i.e.} quantities
that are defined for and depend on an individual configuration of the $N$-particle system.
To distinguish these quantities we will always explicitly state that they depend on time $t$
(like, \textit{e.g.}, in the standard definition of the displacement field mentioned in the 
previous paragraph.) Another example, which will be important in the following, is the microscopic particle density $\rho(\mathbf{r},t)$; it will be precisely defined in Eq.~\eqref{microdensity} below. In crystals, in general the averages of microscopic quantities will change
on the spatial scale of the crystalline cell. For example, the average density in a crystal, $n(\mathbf{r})=\langle \rho(\mathbf{r},t)\rangle$, 
is non-uniform, with large peaks near lattice sites' positions. In contrast, the scalar density, denoted $\delta n(\mathbf{r})$ and the vector
displacement field, $\delta \mathbf{u}(\mathbf{r})$ used in the theory of elasticity vary only on much larger scales; here the $\delta$ indicates a deviation from homogeneous thermal equilibrium. 
 Thus, one
of the goals of Ref.~[\onlinecite{Walz10}] was to identify microscopic fields whose averages correspond 
to the fields of elasticity theory. In the rest of this paper we will call these 
fields microscopic coarse-grained  fields. Also, in the 
rest of the paper we will refer to averages of microscopic quantities as macroscopic fields. Especially, second moments, \textit{viz.} covariances and correlation functions, will be considered in the following and will be connected to thermodynamic derivatives.

\subsection{Microscopic particle density}

The concepts of generalized elasticity theory  \cite{Forster75,Chaikin95} indicate that  density fluctuations close to (all) reciprocal lattice vectors are long-ranged \cite{Wagner66}. Therefore, they all could contribute to coarse-grained fields. 
The microscopic approach to find the displacement field in a real crystal starts from the particle density field $\rho(\mathbf{r},t)$ which depends on the configuration of the $N$-particle system (considering, for simplicity, a one-component crystal of point particles interacting with a spherically symmetric pair-potential) 
 \begin{equation}
\rho(\mathbf{r},t) = \sum_{i=1}^N \delta (\mathbf{r}-\mathbf{r}_i(t))\, \label{microdensity}
\end{equation}
where $\mathbf{r}_i(t)$ are the particle positions, $N$ is the number of particles in the volume $V$; later on we will use $n_0$ to denote the average density, $n_0=N/V$. Spatial Fourier transformation gives fluctuations close to vectors $\bf g$ of the reciprocal lattice 
\begin{equation}
\delta \rho_\mathbf{g}(\mathbf{q},t) = \rho(\mathbf{g+q},t) -  n_{\bf g} V \delta_{\bf q,0}\;, \label{gl2}
\end{equation}
where
\begin{equation}
\rho(\mathbf{k},t) = \int d^d\!r e^{-i\mathbf{k \cdot r}} \rho(\mathbf{r},t)=\sum_i^N e^{-i\mathbf{k\cdot r}_i(t)}\;,
\end{equation}
and
\begin{equation}
n_\mathbf{g} = \frac{1}{V}\langle \rho(\mathbf{g},t)\rangle = \frac{1}{V}\sum_i^N \langle e^{-i\mathbf{g\cdot r}_i(t)}\rangle\;.
\end{equation}
Here,  the general wave vector $\bf k$ was divided up into reciprocal lattice vector $\bf g$ and wave vector $\bf q$, which lies within the first Brillouin zone; $\langle\,\rangle$ brackets indicate canonical averaging at fixed temperature $T$ (averages are time independent due to time-translational invariance of equilibrium states \cite{Hansen96}). 
$n_{\bf g}$ are the Bragg-peak amplitudes (Debye-Waller factors) which serve as crystal order parameters. They quantify the spontaneous breaking of the translational invariance (spatial homogeneity).

\subsection{Coarse-graining microscopic density fluctuations to elasticity fields}

In Ref.~[\onlinecite{Walz10}] the following representation was established for the microscopic density fluctuation in terms of microscopic coarse-grained density and displacement fields
\begin{equation}\label{ansatz}
\delta \rho_\mathbf{g}({\bf q},t) =  -in_\mathbf{g} g_\alpha \delta u_\alpha(\mathbf{q},t) + n_\mathbf{g} \frac{\delta n(\mathbf{q},t)}{n_0}\;, \end{equation}
with Greek indices denoting spatial directions; repeated indices are summed over (Einstein summation convention is used).
Equation (\ref{ansatz}) is the crucial relation linking the fields of macroscopic elasticity theory to the underlying microscopic density fluctuations. It states that for wave vectors $\bf q$  within the first Brillouin zone, the four coarse-grained fields $\delta n(\mathbf{q},t)$ and $\delta \mathbf{u}(\mathbf{q},t)$ determine the hydrodynamic contributions of the microscopic density field. This is valid even close to Bragg-peaks at arbitrarily high reciprocal lattice vectors $\bf g$. Equation \eqref{ansatz} was deduced considering the Zwanzig-Mori equations of motion of the microscopic density fluctuations \cite{Walz10}. In the present contribution, we support  it by considerations of equilibrium correlations.

In ideal crystals without defects the coarse-grained density and the divergence of the displacement field are proportional \cite{Chaikin95}.  In real crystals, mass transport can arise from lattice distortions (described by the displacement field) but also from defect motion, which occurs diffusively over large distances. This additional hydrodynamic mode is called point defect density. It enters  by the standard definition \cite{Chaikin95}:
\begin{align}\label{defect}
\delta c({\bf q},t) = - \delta n({\bf q},t) - i n_0 q_\alpha \delta u_\alpha({\bf q},t)
\; .
\end{align}
In Ref.~[\onlinecite{Walz10}] it is shown that Eqs.~\eqref{ansatz} and \eqref{defect} predict the correct reversible dynamics of  the defect density.  Because many situations require theoretical expressions at constant defect density \cite{Fleming76}, we will use Eq.~\eqref{defect} repeatedly in the following sections.

\subsection{Relating the coarse-grained fields to microscopic density fluctuations}

Explicit expressions for the coarse-grained density and displacement fields can be derived by inverting Eq.~\eqref{ansatz}. The inversion can be performed  using the two summations
\begin{subequations}
\begin{align}
\frac{n_0}{\mathcal{N}_0}\sum_\mathbf{g} n_\mathbf{g}^\ast,\\
\mathcal{N}^{-1}_{\alpha\beta}\sum_\mathbf{g} n_\mathbf{g}^\ast g_\beta\, .
\end{align}\label{summ}
\end{subequations}
and the relation $\sum_\mathbf{g} |n_\mathbf{g}|^2\mathbf{g}=0$. The normalization constants are $\mathcal{N}_0=\sum_\mathbf{g} |n_\mathbf{g}|^2$ and  $\mathcal{N}_{\alpha\beta} = \sum_\mathbf{g} |n_\mathbf{g}|^2 g_\alpha g_\beta$.
Performing the sums over the reciprocal lattice in  Eq.~\eqref{ansatz} leads  to the microscopic coarse-grained density  
\begin{equation}\label{macrodensity}
\delta n(\mathbf{q},t) = \frac{n_0}{\mathcal{N}_0}\sum_\mathbf{g}
n_\mathbf{g}^\ast\; \delta
\rho_\mathbf{g}(\mathbf{q},t)\; ,
\end{equation}
and to the microscopic coarse-grained displacement field
\begin{equation}\label{displacement}
\delta u_\alpha(\mathbf{q},t) = i\mathcal{N}^{-1}_{\alpha\beta}\sum_\mathbf{g} n_\mathbf{g}^\ast\, g_\beta \; \delta \rho_\mathbf{g}(\mathbf{q},t)\;.
\end{equation}
These expressions could be evaluated using information obtained from computer simulations or from colloidal experiments \cite{Reinke07}.

Equations \eqref{macrodensity} and \eqref{displacement} express the coarse-grained fields in terms of microscopic particle density \eqref{microdensity}.
It is intriguing that contributions from all finite lattice vectors ${\bf g}\ne 0$ are present in the coarse-grained density. Even in the limit of vanishing wave vector, $q\to0$, it is not sufficient to  measure particle density fluctuations close to the center of the first Brillouin zone, in order to determine the thermodynamic density field in crystals. Fluctuations from the regions around all lattice vectors contribute and describe how macroscopic strain fluctuations and defect density independently cause changes in the hydrodynamic particle density.    


\section{Relations involving correlations of the coarse-grained fields}

\subsection{Correlation functions of the coarse-grained fields}\label{sect:correlations}

After recalling the relations between the fields of elasticity theory and microscopic fluctuations \cite{Walz10}, we turn now to the focus of our work, the correlations functions of the coarse-grained fields and the thermodynamic derivatives (including the isothermal compressibility). First, we will obtain the correlation functions from classical density functional theory (DFT) \cite{Rowlinson82,Barrat03,Hansen96}. These correlation functions will then be analyzed in the homogeneous case to obtain the thermodynamic quantities.

Close to equilibrium, owing to the fluctuation dissipation theorem, only equilibrium correlation functions are required in order to discuss the linear response to small external fields  \cite{Landau70}. In a homogeneous and unstrained crystal,  the equilibrium correlation functions of the microscopic density fluctuations on the left hand side of Eq.~(\ref{ansatz}) can be calculated within DFT. This enables us to obtain the correlation functions of the coarse-grained fields in Sects.~\ref{sect:correlations}1. and \ref{sect:correlations}2. The fundamental Ornstein-Zernike relation provides a connection between the density correlations and
the inverse density-density correlation matrix $J_{\mathbf{gg^\prime}}(\mathbf{q})$
\begin{align}\label{gl7}
Vk_BT \delta_{\mathbf{g g^{\prime\prime}}} &= \sum_\mathbf{g^\prime} \langle\delta\rho^\ast_\mathbf{g}(\mathbf{q},t)\,\delta\rho_\mathbf{g^\prime}(\mathbf{q},t)\rangle \; J_{\mathbf{g^\prime g^{\prime\prime}}}(\mathbf{q})\;.
\end{align}
Here, the periodicity of the two-point density correlation function \cite{McCarley}
was used
which implies that only density fluctuations whose  wavevectors differ by a vector of the reciprocal lattice are correlated.
The (infinite-dimensional) Hermitian matrix $J_{\mathbf{gg^\prime}}$ is the double Fourier-transform of the second functional derivative of the free energy with respect to the macroscopic density, which includes as non-trivial part the direct correlation function $c({\bf r}_1,{\bf r}_2)$.
\begin{align}\label{Jmatrix}
J_{\mathbf{gg^\prime}}(\mathbf{q}) &=\\ 
 \frac{k_BT}{V}\!\!\! &\int\!\!d^d\!r_1\!\!\!\int\!\!d^d\!r_2 e^{i\mathbf{g\cdot r_1}}e^{-i\mathbf{g^\prime\cdot r_2}} e^{i\mathbf{q\cdot r_{12}}}\!\!\left(\!\frac{\delta(\mathbf{r_{12}})}{n(\mathbf{r_1})}\!-\!c(\mathbf{r_1,r_2})\!\right)\!. \notag 
\end{align}
The direct correlation function $c(\mathbf{r_1,r_2})$ is one of the central quantities of DFT\cite{Rowlinson82,Barrat03}  and is obtained as second functional derivative of the excess free energy $\mathcal{F}^{ex}$ with respect to the average density profile, $k_BT\, c({\bf r}_1,{\bf r}_2)=\frac{\delta^2 \mathcal{F}^{ex}[n(\mathbf{r})]}{\delta n(\mathbf{r_1})\delta n(\mathbf{r_2})}$.  Given an (approximate) expression for the free energy, $J_{\mathbf{gg^\prime}}$  can thus be taken as known. It constitutes the only input for the ensuing theory.
As one consequence, in Sect.~\ref{appFreeEn} below  only the quadratic expression of the free energy functional will play a role and  will be sufficient to evaluate the thermodynamic derivatives required for the elastic response.


\subsubsection{Including coarse-grained density}

It is now conceptually straightforward albeit somewhat tedious to derive the correlation functions of the coarse-grained fields in terms of expressions containing the direct correlation function.  Using Eq.~(\ref{ansatz}), one gets
\begin{align}
 &\langle \delta\rho^\ast_\mathbf{g}(\mathbf{q},t)\delta\rho_\mathbf{g^\prime}(\mathbf{q},t)\rangle = \\
 & n_\mathbf{g}^\ast n_\mathbf{g^\prime}\Big( g_\alpha g^\prime_\beta \langle \delta u^\ast_\alpha(\mathbf{q},t) \delta u_\beta (\mathbf{q},t)\rangle + \langle \frac{\delta n^\ast(\mathbf{q},t) \delta n(\mathbf{q},t)}{n^2_0}\rangle\notag\\
&\quad +i g_\alpha\langle \delta u_\alpha^\ast(\mathbf{q},t)\frac{\delta n(\mathbf{q},t)}{n_0}\rangle -i \langle \frac{\delta n^\ast(\mathbf{q},t)}{n_0} \delta u_\beta(\mathbf{q},t)\rangle g^\prime_\beta \Big)\; , \notag
\end{align}
Inserting this into Eq.~(\ref{gl7}) and with the help of the two summations (\ref{summ}) and Eqs.~\eqref{macrodensity} and \eqref{displacement},
one obtains the following set of equations
\begin{subequations}\label{eq13}
\begin{align}
Vk_BT &= \frac{\langle\delta n^\ast\delta n\rangle}{n_0^2} \nu^\ast(\mathbf{q}) - \langle \frac{\delta n^\ast}{n_0} \delta u_\beta\rangle \mu_\beta(\mathbf{q}),\\
0_\beta &= \frac{\langle\delta n^\ast\delta n\rangle}{n_0^2} \mu^\ast_\beta(\mathbf{q}) - \langle\frac{\delta n^\ast}{n_0} \delta u_\delta\rangle \lambda^\ast_{\delta\beta}(\mathbf{q}),\\
0_\alpha &= \langle \delta u^\ast_\alpha \delta u_\beta\rangle \mu_\beta(\mathbf{q}) - \langle \delta u^\ast_\alpha \frac{\delta n}{n_0}\rangle \nu^\ast(\mathbf{q}),\\
Vk_BT \delta_{\alpha\gamma} &= \langle \delta u^\ast_\alpha \delta u_\beta\rangle \lambda^\ast_{\beta\gamma}(\mathbf{q}) - \langle \delta u^\ast_\alpha \frac{\delta n}{n_0}\rangle \mu^\ast_\gamma(\mathbf{q}).
\end{align}
\end{subequations}
Here, generalized (viz.~$q$-dependent) constants of elasticity,  $\nu(\mathbf{q}), \mu_\alpha(\mathbf{q})$, and $\lambda_{\alpha\beta}(\mathbf{q})$, appear. We will show that they enter into the equilibrium correlation functions of the coarse-grained fields and reduce to thermodynamic derivatives in the long-wavelength limit \cite{Walz10}. Using Eq.~\eqref{Jmatrix}, the $q$-dependent constants of elasticity can be expressed in terms of integrals containing the crystal direct correlation function.
\begin{subequations}\label{numulambda}
\begin{align}
\nu(\mathbf{q}) &= \frac{k_BT}{V}\int d^d\!r_1 \int\! d^d\!r_2 n(\mathbf{r_1}) n(\mathbf{r_2}) e^{-i\mathbf{q\cdot r_{12}}} \notag\\
 &\qquad \times \left(\frac{\delta(\mathbf{r_{12}})}{n(\mathbf{r_1})}-c(\mathbf{r_1,r_2})\right)\\
 &\approx \nu + \mathcal{O}(q^2),\\
\mu_\alpha(\mathbf{q}) &= \frac{k_BT}{V} \int d^d\!r_1 \int d^d\!r_2 c(\mathbf{r_1,r_2})\Big( 1 - e^{-i\mathbf{q\cdot r_{12}}}\Big)\notag\\
 &\qquad \times  n(\mathbf{r_1}) \nabla_\alpha n(\mathbf{r_2}) \\
 &\approx i\mu_{\alpha\beta} q_\beta + \mathcal{O}(q^2),\\
\lambda_{\alpha\beta}(\mathbf{q}) &= \frac{k_BT}{V}\int d^d\!r_1 \int d^d\!r_2 c(\mathbf{r_1,r_2})\Big( 1 - e^{-i\mathbf{q\cdot r_{12}}}\Big) \notag\\
 &\qquad \times \Big(\nabla_\alpha n(\mathbf{r_1})\Big)\Big( \nabla_\beta n(\mathbf{r_2})\Big)  \\
 &\approx \lambda_{\alpha\beta\gamma\delta} q_\gamma q_\delta + \mathcal{O}(q^4).
\end{align}
\end{subequations}
The small wavevector limit and the index-symmetries  $\mu_{\alpha\beta}=\mu_{\beta\alpha}$ and $\lambda_{\alpha\beta\gamma\delta} = \lambda_{\beta\alpha\gamma\delta} = \lambda_{\alpha\beta\delta\gamma} = \lambda_{\gamma\delta\alpha\beta}$ are discussed in detail in Ref.~[\onlinecite{Walz10}]. The explicit integrals are given in Eqs.~\eqref{eq29}, \eqref{eq32} and \eqref{eq34} below, where also crucial steps in their derivation are recalled. The connection of the elastic coefficients to thermodynamic derivatives will be established in Eqs.~\eqref{thermoderiv} and \eqref{thermoderiv2}. 

The obtained set of equations \eqref{eq13} is best presented in matrix notation 
\begin{align}\label{CorrNUMatrix}
V\!k_BT \delta_{ij}\!\! 
&=\!\! \left(\!\begin{array}{c|c}\frac{\langle\delta n^\ast\delta n\rangle}{n_0^2} & -\langle\frac{\delta n^\ast}{n_0}\delta u_\beta \rangle \\ \hline -\langle \delta u^\ast_\alpha\frac{\delta n}{n_0}\rangle & \langle \delta u^\ast_\alpha \delta u_\beta\rangle \end{array}\!\!\right)_{ik}\!\!\!\left(\!\begin{array}{c|c} \nu^\ast(\mathbf{q}) & \mu^\ast_\gamma(\mathbf{q}) \\ \hline \mu_\beta(\mathbf{q})& \lambda^\ast_{\beta\gamma}(\mathbf{q}) \end{array}\!\right)_{kj}\!\!\!,
\end{align}
with Latin indices $i=0,\alpha$.
The matrix of correlation functions of the macroscopic  variables is thus given by the inverse of the matrix of the generalized constants of elasticity
\begin{equation}
\left(\!\begin{array}{c|c}\langle\frac{\delta n^\ast\delta n}{n_0^2}\rangle & -\langle\frac{\delta n^\ast}{n_0}\delta u_\beta\rangle\\\hline -\langle\delta u_\alpha^\ast\frac{\delta n}{n_0}\rangle & \langle\delta u_\alpha^\ast \delta u_\beta\rangle\end{array}\right)\! =\! Vk_BT\! \left(\!\begin{array}{c|c} \nu(\mathbf{q})& \mu^\ast_\beta(\mathbf{q})\\ \hline \mu_\alpha(\mathbf{q})&\lambda_{\alpha\beta}(\mathbf{q})\end{array}\right)^{-1}\!\!.
\end{equation}
The single matrix elements corresponding to the wavevector-dependent correlation functions are\cite{Gantmacher59}
\begin{subequations}\label{InverseAll}
\begin{align}
\langle\frac{\delta n^\ast\delta n}{n_0^2}\rangle &= \!Vk_BT \left(\frac 1\nu + \frac{\mu^\ast_\alpha}{\nu} \Big[\lambda_{\alpha\beta}- \frac{\mu_\alpha\mu_\beta^\ast}{\nu}\Big]^{-1}\frac{\mu_\beta}{\nu}\right)\label{dichtedichte}\\
 &= Vk_BT \left(\nu-\mu^\ast_\alpha \lambda_{\alpha\beta}^{-1}\mu_\beta \right)^{-1}=Vk_BT K^{-1},\\
\langle \delta u_\alpha^\ast\delta u_\beta\rangle &= Vk_BT \left(\lambda_{\alpha\beta}-\mu_\alpha \nu^{-1}\mu_\beta^\ast  \right)^{-1}= Vk_BT H_{\alpha\beta}^{-1}\\
 &= Vk_BT \left(\lambda_{\alpha\beta}^{-1} + \lambda_{\alpha\gamma}^{-1}\mu_\gamma K^{-1} \mu_\delta^\ast \lambda_{\delta\beta}^{-1} \right),\\
-\langle\frac{\delta n^\ast}{n_0}\delta u_\beta\rangle &= Vk_BT \left( -\nu^{-1}\mu_\alpha^\ast H_{\alpha\beta}^{-1}\right)\\
 &= Vk_BT\left( -K^{-1}\mu_\alpha^\ast\lambda_{\alpha\beta}^{-1}\right),\\
-\langle \delta u_\alpha^\ast \frac{\delta n}{n_0}\rangle &= Vk_BT\left( -H_{\alpha\beta}^{-1}\mu_\beta \nu^{-1}\right)\\*
 &= Vk_BT \left(-\lambda_{\alpha\beta}^{-1} \mu_\beta K^{-1}\right).
\end{align}
\end{subequations}
The second line of each expression is a non-trivial alternative, which is here given for completeness; it is based on the algebraic Woodbury identity. 

We thus reached our first goal of expressing the correlation functions of the coarse-grained variables, hydrodynamic density and displacement vector field, in terms of integrals containing the functional derivative of the free energy with respect to density, \textit{viz}.~the direct correlation function. 
Let us note in passing that translational symmetry \cite{Walz10} yields the expected $q$-divergences or $q$-dependences of the correlation functions in the limit $q\to 0$. In particular, $\langle \delta u_\alpha^\ast\delta u^{ }_\beta\rangle \propto 1/q^{2}$ follows from $\lambda_{\alpha\beta}({\bf q})\propto q^2$. 

\subsubsection{Including defect density}

Although the relation between the constants of elasticity and the fluctuations of the coarse-grained fields is complete, it is worthwhile to consider a second set of variables. So far the displacement field $u_\alpha$ appeared in two different ways. It contributes to the coarse-grained density, but it also appears as broken symmetry variable. In this section we introduce the point defect density $c$ in lieu of the coarse-grained density, and keep the displacement field.

The correlation functions of the coarse-grained density and displacement field are easily transformed into correlations between the fluctuations of the point defect density and the displacement field  using the definition Eq.~\eqref{defect}.
The set of variables $\{\delta c(\mathbf{q}), \delta u_\alpha(\mathbf{q})\}$ may be more appropriate to describe an experiment when few defects are present and $\delta c(\mathbf{q},t)$ can be measured easily. It allows one to take the limit of vanishing defect density and thus it is a natural set of variables to be used when defects are neglected. Thus, it correctly captures the ideal crystal limit. Eq. \eqref{CorrNUMatrix} is transformed into
\begin{align}
& Vk_BT \delta_{ij} = & \nonumber \\
& \left(\begin{array}{c|c}\frac{\langle\delta c^\ast\delta c\rangle}{n_0^2} & \langle\frac{\delta c^\ast}{n_0} \delta u_\alpha \rangle \\ \hline \langle \delta u^\ast_\sigma \frac{\delta c}{n_0}\rangle & \langle \delta u^\ast_\sigma \delta u_\alpha \rangle \end{array}\right)_{ik} \left(\begin{array}{c|c} \nu^\ast(\mathbf{q}) & n_0 V_\delta({\bf q}) \\ \hline n_0 V^\ast_\alpha({\bf q})& \Lambda^\ast_{\alpha\delta}({\bf q}) \end{array}\right)_{kj}. &  \label{CorrNUMatrix2}
\end{align}
The combination of the constants of elasticity appearing here is directly connected to  the hydrodynamic equation of the momentum density expressed in terms of point defect density and displacement field as hydrodynamic variables\cite{Walz10}. There, the time derivative of the momentum density couples to the displacement field via the negative of
\begin{equation}
\Lambda_{\alpha\beta}(\mathbf{q}) = \lambda_{\alpha\beta}(\mathbf{q}) - i q_\alpha \mu_\beta(\mathbf{q}) + i\mu^\ast_\alpha(\mathbf{q}) q_\beta + q_\alpha \nu(\mathbf{q}) q_\beta.\label{eq19}
\end{equation}
The coupling to the point defect density is given by the negative of
\begin{equation}
V_\alpha(\mathbf{q}) = \frac{1}{n_0} \Big(\mu^\ast_\alpha(\mathbf{q}) - iq_\alpha \nu(\mathbf{q})\Big).
\end{equation}
The individual matrix elements of the correlation functions in terms of $\nu(\mathbf{q})$, $V_\alpha(\mathbf{q})$, and $\Lambda_{\alpha\beta}(\mathbf{q})$, and the limit $q\to 0$, may be determined according to the steps in the previous  paragraphs. As the results can be obtained from Eqs.~\eqref{InverseAll} by straightforward replacements, identified from comparing Eqs.~\eqref{CorrNUMatrix}  and \eqref{CorrNUMatrix2}, they will not be repeated here.

\subsection{Inverse relations}

Equations (\ref{InverseAll}) predict the fluctuations of the macroscopic coarse-grained density and displacement field based on the generalized constants of elasticity obtained from the direct correlation function and thus the free energy. Experimentally, the inverse relations are of interest: expressing the generalized constants of elasticity of the crystal in terms of measurable correlation functions. Two different sets of correlation functions can be obtained from experiments. Either displacement field and coarse-grained density fluctuations can be recorded,  or displacement field and defect density. For reference, we provide the inverse relations for both cases in this section.

\subsubsection{Including coarse-grained density}

The matrix equation (\ref{CorrNUMatrix}) can be inverted in order to find the elastic functions $\nu({\bf q})$, $\mu_\alpha({\bf q})$, and $\lambda_{\alpha\beta}({\bf q})$ in terms of measurable fluctuation functions. The inverse relations read:
\begin{widetext}
\begin{subequations}\label{inverse}
\begin{align}
\frac{\nu(\mathbf{q})}{Vk_BT} &= \langle\frac{\delta n^\ast\delta n}{n_0^2}\rangle^{-1} + \langle\frac{\delta n^\ast\delta n}{n_0^2}\rangle^{-1} \langle\frac{\delta n^\ast}{n_0}\delta u_\alpha\rangle \Big[\langle\delta u^\ast_\alpha \delta u_\beta\rangle -\langle\delta u^\ast_\alpha\frac{\delta n}{n_0}\rangle \langle\frac{\delta n^\ast\delta n}{n_0^2}\rangle^{-1} \langle\frac{\delta n^\ast}{n_0}\delta u_\beta\rangle \Big]^{-1} \langle\delta u^\ast_\beta\frac{\delta n}{n_0}\rangle \langle\frac{\delta n^\ast\delta n}{n_0^2}\rangle^{-1}\\
 &= \Big( \langle\frac{\delta n^\ast\delta n}{n_0^2}\rangle - \langle\frac{\delta n^\ast}{n_0}\delta u_\alpha\rangle \langle\delta u^\ast_\alpha \delta u_\beta\rangle^{-1} \langle \delta u^\ast_\beta\frac{\delta n}{n_0}\rangle  \Big)^{-1} = R^{-1},\\
\frac{\lambda_{\alpha\beta}(\mathbf{q})}{Vk_BT} &= \Big( \langle\delta u^\ast_\alpha \delta u_\beta\rangle -\langle\delta u^\ast_\alpha\frac{\delta n}{n_0}\rangle \langle\frac{\delta n^\ast\delta n}{n_0^2}\rangle^{-1} \langle\frac{\delta n^\ast}{n_0}\delta u_\beta\rangle  \Big)^{-1} = S^{-1}_{\alpha\beta}\\
 &= \langle\delta u^\ast_\alpha \delta u_\beta\rangle^{-1} + \langle \delta u^\ast_\alpha \delta u_\gamma\rangle^{-1} \langle\delta u^\ast_\gamma \frac{\delta n}{n_0}\rangle R^{-1} \langle\frac{\delta n^\ast}{n_0}\delta u_\delta\rangle \langle\delta u^\ast_\delta \delta u_\beta\rangle^{-1},\\
\frac{\mu_\alpha(\mathbf{q})}{Vk_BT} &= S^{-1}_{\alpha\beta} \langle\delta u^\ast_\beta\frac{\delta n}{n_0}\rangle \langle\frac{\delta n^\ast\delta n}{n_0^2}\rangle^{-1}\\
 &= \langle\delta u^\ast_\alpha \delta u_\beta\rangle^{-1} \langle\delta u^\ast_\beta\frac{\delta n}{n_0}\rangle R^{-1}.
\end{align}
\end{subequations}
\end{widetext}
We thus reached our second goal to derive relations which determine the generalized elasticity constants $\lambda_{\alpha\beta}(\mathbf{q})$, $\mu_\alpha(\mathbf{q})$, and $\nu(\mathbf{q})$ from measurements of correlation functions.

\subsubsection{Including defect density}\label{Kap3B}

Replacing the total density with the defect density using Eq.~\eqref{defect}, the generalized constants of elasticity can be connected to fluctuation functions which can be measured at constant (possibly vanishing) defect density. The comparison of the matrices in Eqs.~\eqref{CorrNUMatrix} and \eqref{CorrNUMatrix2} indicates the straightforward replacements in Eq.~\eqref{inverse}. The dynamical matrix $\Lambda_{\alpha\beta}(\mathbf{q})$ determines the wave equation of the momentum density, and its eigenvalues give  the (acoustic) phonon dispersion relations. Because the results follow from straightforward replacements, they will not be given explicitly here.

\section{Free energy and thermodynamic relations}
\label{appFreeEn}

In order to obtain the thermodynamics derivatives, a consideration of the free energy appears useful in cases where the connection to the small wavevector limit of the  correlation functions is not established or under debate \cite{Stillinger66,Goetze67,Goetze68}. In this section, we will coarse-grain the free energy functional of density functional theory in order to derive the thermodynamic derivatives. This purely equilibrium statistical mechanics approach  supplements the dynamical one based on projection operator formalism in Ref.~[\onlinecite{Walz10}].  Importantly, the wavevector dependent correlation functions of the coarse-grained fields of elasticity theory and the thermodynamic elastic free energy  of real (viz.~defect containing) crystals are then obtained from a single microscopic starting point. 

\subsection{Coarse-grained free energy functional with elastic fields}

The second order change in free energy $\Delta \mathcal{F}$ due to a deviation $\delta \rho(\mathbf{r})$ in the average  density distribution  from the periodic crystalline equilibrium density $n(\mathbf{r})$ is \cite{Hansen96,Evans79,Rowlinson82}
\begin{align}
\Delta \mathcal{F} &= \frac{k_BT}{2}\int\!\!\int d^d\!r_1 d^d\!r_2 \Big(\frac{\delta(\mathbf{r_{12}})}{n(\mathbf{r_1})}-c(\mathbf{r_1,r_2})\Big) \delta \rho(\mathbf{r_1})\delta\rho(\mathbf{r_2}),\label{FreeEnergyDFT}
\end{align}
where $c(\mathbf{r_1,r_2})$ is the direct correlation function of a periodic crystal.
Note that this quadratic functional contains the direct correlation function as single input and thus the identical information as used in the correlation functions approach of the previous Sect.~\ref{sectII}. 

\subsubsection{Including coarse-grained density}

We start from the representation of the microscopic density fluctuation in terms of 
displacement field and coarse-grained density, Eq.~\eqref{ansatz}. We assume that an
analogous equation holds also for the averaged (macroscopic) densities. In this way
we get a change of the average density due to non-vanishing displacement field and
average coarse-grained density,
\begin{align}
\delta\rho(\mathbf{r}) &= -\delta\mathbf{u}(\mathbf{r})\cdot\mathbf{\nabla}n(\mathbf{r}) + n(\mathbf{r})\frac{\delta n(\mathbf{r})}{n_0},
\label{relationdnucave}
\end{align}
We shall emphasize that while $\delta\rho(\mathbf{r})$ varies on the spatial scale of the
crystalline lattice, the coarse-grained density  varies far more smoothly and contains 
wavevector contributions only from the first Brillouin zone:
\begin{equation}
\delta n(\mathbf{r}) = \int_{\rm 1^{st}\,BZ} \frac{d^d{q}}{(2\pi)^d}\; e^{i \mathbf{q} \cdot \mathbf{r}}\; \delta n(\mathbf{q})\; . \notag
\end{equation}

Using Eq.~(\ref{relationdnucave}) we obtain the following expression for the product of density changes
\begin{align}
 &\delta \rho(\mathbf{r_1})\delta\rho(\mathbf{r_2}) = \!\underbrace{\delta u_\alpha(\mathbf{r_1}) \delta u_\beta(\mathbf{r_2})\nabla_\alpha n(\mathbf{r_1}) \nabla_\beta n(\mathbf{r_2})}_{(1.)} \notag\\
 &+ \underbrace{\frac{n(\mathbf{r_1})n(\mathbf{r_2}) \delta n(\mathbf{r_1}) \delta n(\mathbf{r_2})}{n^2_0}}_{(2.)} \underbrace{- \delta u_\alpha(\mathbf{r_1})\!\nabla_\alpha n(\mathbf{r_1}) \frac{n(\mathbf{r_2})\delta n(\mathbf{r_2})}{n_0}}_{(3.)} \notag\\
 &\underbrace{- \frac{n(\mathbf{r_1})\delta n(\mathbf{r_1})}{n_0} \delta u_\alpha(\mathbf{r_2})\!\nabla_\alpha n(\mathbf{r_2})}_{(4.)}.\label{drhosquared}
\end{align}
In the following, we substitute the four parts of Eq.~(\ref{drhosquared}) into Eq.~\eqref{FreeEnergyDFT}. We denote the resulting expressions $\Delta\mathcal{F}_{(i.)}$, where $i=1, ..., 4$. We then re-write these expressions using the LMB\cite{Lovett76}W\cite{Wertheim76} equation 
\begin{align}\label{LMBW1}
\frac{\nabla_\alpha n(\mathbf{r})}{n(\mathbf{r})} &= \int d^d\!r^\prime c(\mathbf{r,r^\prime})\nabla_\alpha n(\mathbf{r^\prime}).
\end{align}
Our subsequent calculation is analogous to that of Masters \cite{Masters01} and is equivalent to the discussion of the surface tension in [\onlinecite{Schofield82}]. We will in detail describe the calculation originating from the first part of Eq.~(\ref{drhosquared}), which leads to the elastic tensor $\lambda$, and then summarize calculations originating from the other parts.

In the expression for $\Delta\mathcal{F}_{(1.)}$ one expands $\delta u_\beta(\mathbf{r_2})$ around $\mathbf{r_1}$, which is valid for a short range (in $\mathbf{r_{12}}$) direct correlation function. The zero order term vanishes, because of \eqref{LMBW1} and the first order term does not contribute due to the symmetry $c(\mathbf{r_1,r_2})=c(\mathbf{r_2,r_1})$. Since the hydrodynamic variable $\delta \mathbf{u(r)}$ is slowly varying, one obtains 
an expression which is quadratic in $\nabla\delta \mathbf{u(r)}$ as leading contribution
\begin{widetext}
\begin{align}
\Delta\mathcal{F}_{(1.)}\! &= \frac{k_BT}{2}\int\!\! \int d^d\!r_1 d^d\!r_2 \Big(\frac{\delta(\mathbf{r_{12}})}{n(\mathbf{r_1})}-c(\mathbf{r_1,r_2})\Big) \delta u_\alpha(\mathbf{r_1})\delta u_\beta(\mathbf{r_2}) \nabla_\alpha n(\mathbf{r_1})\nabla_\beta n(\mathbf{r_2}) \label{eq26} \\
 &=\! \frac{k_BT}{2}\!\!\int\!\!\! \int\! d^d\!r_1 d^d\!r_2 \nabla_\alpha n(\mathbf{r_1})c(\mathbf{r_1,r_2})\nabla_\beta n(\mathbf{r_2}) \delta u_\alpha(\mathbf{r_1})
\Big(\delta u_\beta(\mathbf{r_1})\!-\!\delta u_\beta(\mathbf{r_1})\!+\underbrace{\nabla_\gamma \delta u_\beta(\mathbf{r_1}) r_{12,\gamma}}_{=0\text{ symmetry}} - \frac{1}{2}\nabla_\gamma\nabla_\delta \delta u_\beta(\mathbf{r_1}) r_{12,\gamma}r_{12,\delta}\Big) \notag \\&=\! \frac{1}{2}\!\!\int\!\!\! \int\! d^d\!r_1 d^d\!r_2 \left[ \frac{k_BT}{2} \nabla_\alpha n(\mathbf{r_1})c(\mathbf{r_1,r_2})\nabla_\beta n(\mathbf{r_2})  r_{12,\gamma}r_{12,\delta} \right]  \nabla_\gamma \delta u_\alpha(\mathbf{r_1})\nabla_\delta \delta u_\beta(\mathbf{r_1}) \notag \\
 &= \frac{1}{2}\int d^d\!r \lambda_{\alpha\beta\gamma\delta} \nabla_\gamma \delta u_\alpha(\mathbf{r}) \nabla_\delta \delta u_\beta(\mathbf{r}),\notag \\
\lambda_{\alpha\beta\gamma\delta} &= \frac{k_BT}{2V}\int\!\!\int d^d\!r_1 d^d\!r_2 \nabla_\alpha n(\mathbf{r_1})c(\mathbf{r_1,r_2})\nabla_\beta n(\mathbf{r_2}) r_{12,\gamma}r_{12,\delta} \; .\label{eq29}
\end{align}
\end{widetext}
In the last line of Eq.~\eqref{eq26} the separation of spatial scales was used in order to replace the slowly varying local elastic coefficient given by the contents of the square bracket on the third line of  Eq.~\eqref{eq26} by the  macroscopic constant of elasticity $\lambda_{\alpha\beta\gamma\delta}$. We emphasize that the expression \eqref{eq29} agrees with the one obtained in the framework of hydrodynamic equations of motion\cite{Walz10}, which was reproduced in Eq.~(\ref{numulambda}f). 

For the second term of the free energy, $\delta n(\mathbf{r_2})$ is expanded around $\mathbf{r_1}$ and, as hydrodynamic variable, assumed to be slowly varying
\begin{align}
\Delta\mathcal{F}_{(2.)} &=\!\frac{k_BT}{2}\!\!\int\!\!\int\! d^d\!r_1 d^d\!r_2 \frac{\delta n(\mathbf{r_1})\delta n(\mathbf{r_2})}{n^2_0} \notag\\
 & \qquad\qquad \times [n(\mathbf{r_1})\delta(\mathbf{r_{12}})\!-\!n(\mathbf{r_1})c(\mathbf{r_1,r_2})n(\mathbf{r_2})]  ,\\
 &= \frac{1}{2}\int d^d\!r \ \nu \ \Big(\frac{\delta n(\mathbf{r})}{n_0}\Big)^2.
\end{align}
With
\begin{align}
\nu &= \frac{k_BT}{V}\!\!\int\!\!\!\int\!\! d^d\!r_1 d^d\!r_2 \Big(n(\mathbf{r_1})\delta(\mathbf{r_{12}})\!-n(\mathbf{r_1})c(\mathbf{r_1,r_2})n(\mathbf{r_2})\Big). \label{eq32}
\end{align}
The third and fourth part yield with the same arguments
\begin{align}
\Delta\mathcal{F}_{(3.+4.)}\! 
 &= -\int d^d\!r \ \mu_{\alpha\beta}\ \frac{\delta n(\mathbf{r})}{n_0} \nabla_\beta \delta u_\alpha(\mathbf{r}),\\
\mu_{\alpha\beta}\! &=\! \frac{k_BT}{V}\!\int\! d^d\!r_1\! \int d^d\!r_2 n(\mathbf{r_1}) \nabla_\alpha n(\mathbf{r_2}) r_{12,\beta} c(\mathbf{r_1,r_2})\label{eq34}
\end{align}

Summarizing, we obtain the following expression for the free energy change
\begin{align}\label{FreeEnN}
\Delta\mathcal{F} &= \frac{1}{2}\int d^d\!r \ \nu \Big(\frac{\delta n(\mathbf{r})}{n_0}\Big)^2 + C^n_{\alpha\beta\gamma\delta}  u_{\alpha\beta}(\mathbf{r})  u_{\gamma\delta}(\mathbf{r})\notag \\
 &\quad - \int d^d\!r  \mu_{\alpha\beta}\frac{\delta n(\mathbf{r})}{n_0}  u_{\alpha\beta}(\mathbf{r})
\end{align}
Expression (\ref{FreeEnN}) involves the symmetrized linear strain tensor $u_{\alpha\beta}(\mathbf{r})=\frac{1}{2}(\nabla_\alpha \delta u_\beta(\mathbf{r}) + \nabla_\beta \delta u_\alpha(\mathbf{r}))$ and the Voigt-symmetric elastic constants  $C^n_{\alpha\beta\gamma\delta}=\lambda_{\alpha\gamma\beta\delta} + \lambda_{\beta\gamma\alpha\delta} - \lambda_{\alpha\beta\gamma\delta}$. Both combinations reflect the rotational symmetry as only symmetric combinations of strain enter into the elastic energy and the tensor of elastic constants obeys a number of symmetry relations. Their proof \cite{Walz10} is based upon the rotational analog of the LMBW equation\cite{Schofield82}
\begin{align}
\mathbf{r_1}\!\times\!\nabla^{(1)} \ln n(\mathbf{r_1}) &=\! \int\! d^d\!r_2 c(\mathbf{r_1,r_2})\Big(\!\mathbf{r_2}\!\times\! \nabla^{(2)}n(\mathbf{r_2})\!\Big)\!.
\end{align}

We thus arrived at our third goal, to derive the general elastic free energy functional of real crystals containing the coarse-grained macroscopic fields.  Let us add that the above expression for the free energy also determines the constant $C_0 =0$ in Eqs. (89), (90), and (92) of Ref.~[\onlinecite{Walz10}], which could not be determined from the hydrodynamic equations considered there.

\subsubsection{Including defect density}

The free energy in terms of the defect density $\delta c(\mathbf{r})$ and the displacement field $\delta \mathbf{u(r)}$ is obtained from Fourier transforming ansatz \eqref{ansatz} and Eq.~\eqref{defect}  into real space:
\begin{equation}\label{ansatz2b}
\delta\rho(\mathbf{r},t) = -\nabla\cdot \left[  n(\mathbf{r}) \delta \mathbf{u}(\mathbf{r},t)  \right]- \frac{n(\mathbf{r})}{n_0} \delta c(\mathbf{r},t) \, . 
\end{equation}
Following the steps of the previous section one arrives at the coarse-grained free energy including the defect density:
\begin{align}\label{FreeEnC}
& \Delta\mathcal{F} =\!\frac{1}{2}\!\int\! d^d\!r  \nu \Big(\frac{\delta c(\mathbf{r})}{n_0}\Big)^2\! 
 + 2\Big( \nu \delta_{\alpha\beta} +\mu_{\alpha\beta}\Big) \frac{\delta c(\mathbf{r})}{n_0} u_{\alpha\beta}(\mathbf{r}) \notag\\
&+\!\Big(C^n_{\alpha\beta\gamma\delta}\!+\!\nu\delta_{\alpha\beta}\delta_{\gamma\delta}\!+\!\mu_{\alpha\beta}\delta_{\gamma\delta}\!+\!\delta_{\alpha\beta}\mu_{\gamma\delta} \Big) u_{\alpha\beta}(\mathbf{r})  u_{\gamma\delta}(\mathbf{r}).
\end{align}
This gives the relation between the elastic coefficients at given defect density $C^c$ in terms of the corresponding coefficients at given total density, $C^n$, namely:
$C^c_{\alpha\beta\gamma\delta}=C^n_{\alpha\beta\gamma\delta}\!+\!\nu\delta_{\alpha\beta}\delta_{\gamma\delta}\!+\!\mu_{\alpha\beta}\delta_{\gamma\delta}\!+\!\delta_{\alpha\beta}\mu_{\gamma\delta}$.

\subsubsection{Gaussian probability distribution function}

The harmonic free energy Eq.~\eqref{FreeEnN} can be written in a more compact form with the help of the $4\times4$-matrix of elastic coefficients introduced in Eq.~\eqref{CorrNUMatrix}. Fourier-transformation leads to
\begin{align}
\Delta\mathcal{F} &= \frac{1}{2}\int \frac{d^d\!q}{(2\pi)^d} \\
& \left(\!\!\begin{array}{cc}\frac{\delta n^\ast({\bf q})}{n_0},& \delta u^\ast_\alpha({\bf q}) \end{array}\!\!\right)\left(\begin{array}{cc} \nu & -i \mu_{\gamma\delta} q_\delta \\ i \mu_{\alpha\beta} q_\beta & C^n_{\alpha\beta\gamma\delta}q_\beta q_\delta \end{array}\right)\left( \!\begin{array}{c} \frac{\delta n({\bf q})}{n_0}\\ \delta u_\gamma({\bf q})\end{array}\!\right)\notag
\end{align}
This free energy functional is a superposition of independent terms each containing the square of the Fourier transformed coarse-grained fields at one specific wavevector. Often one connects such quadratic free energy functionals with a probability distribution for fluctuations of the coarse-grained fields \cite{Chaikin95}; $P[\delta n(\mathbf{q}),\delta \mathbf{u}(\mathbf{q})]\propto \exp{\{-\Delta\mathcal{F}/k_BT\}}$. In the present case, this would yield the  wavevector-dependent correlation functions \eqref{CorrNUMatrix} as statement of the equipartition theorem resulting from this Gaussian approximation.

\subsection{The thermodynamic elastic free energy}

The thermodynamic free energy corresponds to homogeneous fluctuations, viz.~the coarse-grained fields evaluated at ${\bf q}=0$. It can handily be obtained from the explicit  free energy functional in Eq.~\eqref{FreeEnN}. The result shall be given using 
the Voigt notation \cite{Ashcroft76} (in three dimensions), because this form appears convenient for explicit model calculations later on. Quantities in Voigt notation  carry lower Latin indices $1\le i\le 6$, where $u_i$ denotes the independent elements of the symmetric strain tensors  $u_{\alpha\beta}$.  For $1\le i \le 3$ the relation $u_i=u_{\alpha,\beta}$ holds with $(\alpha,\beta)=\{(1,1);(2,2);(3,3)\}$, while for  $4\le i \le 6$,  $u_i=2 u_{\alpha,\beta}$ holds with $(\alpha,\beta)=\{(2,3);(1,3);(1,2)\}$. For spatially constant fluctuations (to be indicated by subscript $q=0$ where otherwise unclear), one obtains in obvious notation as a quadratic form:
\begin{align}
\Delta\mathcal{F} &= \frac{V}{2} \left( \begin{array}{cc}\frac{\delta n}{n_0},& u_i \end{array}\right)\left(\begin{array}{cc} \nu & -\mu_j\\ -\mu_i& C^n_{ij}\end{array}\right)\left( \begin{array}{c} \frac{\delta n}{n_0}\\ u_j\end{array}\right)\label{FreeEnThermo} 
\end{align}
The thermodynamic free energy is a quadratic form given by a $7\times7$-matrix of elastic coefficients, where in Voigt notation the elastic matrix is $C_{ij}=C_{\alpha\beta\gamma\delta}$ for $1\le i,j \le 6$ with the index correspondences given above.

\subsubsection{Connection to elastic coefficients and variances}

Thermodynamic derivatives can now easily be evaluated and lead to the parameters already introduced in Eq.~\eqref{numulambda}. 
They follow from the Gibbs fundamental form of the free energy density $f=F/V\approx\Delta\mathcal{F}/V$, where the  quadratic expression \eqref{FreeEnThermo} suffices in order to obtain the second order derivatives of interest.  
\begin{subequations}\label{thermoderiv}
\begin{align}
\frac{\partial^2 f}{\partial n^2}\Big|_{u_{\alpha\beta}} &= \frac{\partial \mu}{\partial n}\Big|_{u_{\alpha\beta}}= \nu /n_0^2 ,\\
\frac{\partial^2 f}{\partial n \partial u_{\alpha\beta}} &= \frac{\partial \mu}{\partial u_{\alpha\beta}}\Big|_{n} =
\frac{\partial h_{\alpha\beta}}{\partial n}\Big|_{u_{\gamma\delta}} =
-\mu_{\alpha\beta} / n_0,\\
\frac{\partial^2 f}{\partial u_{\alpha\beta} \partial u_{\gamma\delta}}\Big|_{n} &= \frac{\partial h_{\alpha\beta}}{\partial u_{\gamma\delta}}\Big|_{n} = C^n_{\alpha\beta\gamma\delta}= \!\lambda_{\alpha\gamma\beta\delta}\! +\! \lambda_{\beta\gamma\alpha\delta}\! -\! \lambda_{\alpha\beta\gamma\delta}.
\end{align}
\end{subequations}
These relations identify the elastic parameters of our approach as thermodynamic derivatives. They already use the familiar intensive variables, chemical potential $\mu$ and stress tensor $h_{\alpha\beta}$ in order to familiarize with later relations \cite{Chaikin95,Fuchs2012}. These variables will be introduced and discussed in Sect.~\ref{isothermal} below.
Let us note that these calculations supplement the derivation of the thermodynamic relations in Ref.~[\onlinecite{Walz10}] (recalled in Eq.~\eqref{numulambda}), where the equivalence of the hydrodynamic equations was used. The thermodynamic free energy thus takes the form:
\begin{align}
\Delta\mathcal{F} 
&=\frac{V}{2} \left( \begin{array}{cc}\frac{\delta n}{n_0},& u_i \end{array}\right)\left(\begin{array}{cc} 
n_0^2\frac{\partial \mu}{\partial n} & -n_0\frac{\partial \mu}{\partial u_{j}}\\ -n_0\frac{\partial h_{i}}{\partial n} &\frac{\partial h_{i}}{\partial u_{j}}\end{array}\right)\left( \begin{array}{c} \frac{\delta n}{n_0}\\ u_j\end{array}\right) 
\end{align}
Where, in Voigt notation the stresses correspond  to $h_i=h_{\alpha\beta}$ for $1\le i \le 6$.

This compact expression is a convenient starting point for evaluating the thermodynamic covariances and susceptibilities which enter elasticity theory. The isothermal compressibility and the defect density susceptibility will be obtained in the next Sect.~\ref{isothermal}. In order to prepare for this, first the second moments of the fluctuations of the thermodynamic variables shall be obtained. These are connected to the thermodynamic derivatives using the thermodynamic formalism. 
Because the inverse of the Jacobian matrix is equal to the Jacobian matrix of the inverse function one obtains
\begin{align}\label{eq42}
\left(\begin{array}{cc} \nu & -\mu_j\\ -\mu_i& C^n_{ij}\end{array}\right)^{-1} 
&=\left(\begin{array}{cc} 
n_0^2\frac{\partial \mu}{\partial n} & -n_0\frac{\partial \mu}{\partial u_{j}}\\ -n_0\frac{\partial h_{i}}{\partial n} &\frac{\partial h_{i}}{\partial u_{j}}\end{array}\right)^{-1} \\ \notag
&=\left(\begin{array}{cc} 
\frac{1}{n_0^2}\frac{\partial n}{\partial \mu} & -\frac{1}{n_0}\frac{\partial  u_{j}}{\partial \mu}\\ -\frac{1}{n_0}\frac{\partial n}{\partial h_{i}} &\frac{\partial u_{j}}{\partial h_{i}}\end{array}\right)\\\notag
&=\frac{1}{Vk_BT}\left(\begin{array}{cc} \frac{\langle \delta n\delta n\rangle}{n_0^2} & \langle \frac{\delta n}{n_0} u_j \rangle \\ \langle u_i \frac{\delta n}{n_0}\rangle & \langle u_i u_j\rangle \end{array} \right)\Big|_{q=0}
\end{align}
In the last step the fluctuation-dissipation-theorem is used\cite{Forster75}. The variance  of the total coarse-grained density variation is thus obtained from a simple matrix inversion \cite{Gantmacher59}:
\begin{align}\label{eq43}
\langle \frac{\delta n \delta n}{n^2_0}\rangle\Big|_{q=0} &= Vk_BT \Big(\frac{1}{\nu} + \frac{\mu_i}{\nu}\Big[ C^n_{ij} -\frac{\mu_i \mu_j}{\nu}\Big]^{-1} \frac{\mu_j}{\nu} \Big)\\\notag
&=Vk_BT \Big(\nu-\mu_i (C^n_{ij})^{-1} \mu_j \Big)^{-1}\\
&= Vk_BT \Big( \nu \! -\! \mu_{\alpha\beta} \Big[ C^n_{\alpha\beta\gamma\delta} \Big]^{-1}\! \mu_{\gamma\delta} \Big)^{-1}\notag,
\end{align}
where the second line follows from a Woodbury identity, and  the usual notation is used instead of the Voigt one in the last line; see Wallace \cite{Wallace70} and the Appendix \ref{appComp} for the proper interpretation of the inverse fourth-rank tensor. 

We thus derived the second moment of the particle number fluctuations from DFT. We started from the same free energy functional as was used in the derivation of the wavevector-dependent correlation functions summarized in Eq.~\eqref{InverseAll}. Thus, in Sect.~\ref{sect:small}, both results can be compared in the long-wavelength limit.

\subsubsection{Including defect density}

In a similar manner an expression for the defect density fluctuation can be obtained. Starting from the free energy functional in Eq.~\eqref{FreeEnC} and considering homogeneous variations (viz.~ at $\mathbf q=0$), one recognizes that the relevant thermodynamic derivatives are now given by
\begin{subequations}
\begin{align}
\frac{\partial^2 f}{\partial c^2}\Big|_{u_{\alpha\beta}} &= -\frac{\partial \mu}{\partial c}\Big|_{u_{\alpha\beta}} = \nu/ n_0^2,\\
\frac{\partial^2 f}{\partial c\partial u_{\alpha\beta}} &= - \frac{\partial \mu}{\partial u_{\alpha\beta}}\Big|_{c} = \frac{\partial \sigma_{\alpha\beta}}{\partial c}\Big|_{u_{\alpha\beta}} \nonumber\\ &= \left(\nu\delta_{\alpha\beta}+\mu_{\alpha\beta}\right)/n_0 =\mu_{\alpha\beta}^c/n_0,\label{eq44b}\\
\frac{\partial^2 f}{\partial u_{\alpha\beta}\partial u_{\gamma\delta}}\Big|_{c}\!\! &=\! C^c_{\alpha\beta\gamma\delta}\! = \!C^n_{\alpha\beta\gamma\delta}\! +\!\mu_{\alpha\beta}\delta_{\gamma\delta} \!+\! \delta_{\alpha\beta}\mu_{\gamma\delta}\! +\! \nu\delta_{\alpha\beta}\delta_{\gamma\delta},\label{ElastKonstCc}
\end{align}\label{thermoderiv2}
\end{subequations}
where the stress tensor $\sigma_{\alpha\beta}$ was introduced, which will be discussed in in Sect.~\ref{isothermal} below. Also the abbreviation ${\boldsymbol \mu}^c$ was introduced.
Thus, using the fluctuation dissipation theorem again,  the
matrix of total thermodynamic variations is given by
\begin{align}\label{Matrix_c}
\left(\begin{array}{cc} \frac{\langle \delta c\delta c\rangle}{n_0^2} & \langle \frac{\delta c}{n_0} u_j \rangle \\ \langle u_i \frac{\delta c}{n_0}\rangle & \langle u_i u_j\rangle \end{array} \right)\Big|_{q=0} \nonumber
&=Vk_BT\left(\begin{array}{cc} 
\frac{-1}{n_0^2}\frac{\partial c}{\partial \mu} & -\frac{1}{n_0}\frac{\partial  u_{j}}{\partial \mu}\\ \frac{1}{n_0}\frac{\partial c}{\partial \sigma_{i}} &\frac{\partial u_{j}}{\partial \sigma_{i}}\end{array}\right)\\ 
&= Vk_BT \left(\begin{array}{cc} \nu & \mu^c_j\\ \mu^c_i& C^c_{ij}\end{array}\right)^{-1}
\end{align}
$C^c_{ij}$ and $\mu_i^c$ are the tensors from Eq.~\eqref{ElastKonstCc} and Eq.~\eqref{eq44b} in Voigt notation. This leads to the correlation of the 
defect density fluctuations
\begin{align}\label{kappac}
& \langle \frac{\delta c \delta c}{Vk_BTn^2_0}\rangle\Big|_{q=0} 
=\Big(\frac{1}{\nu} + \frac{\mu^c_i}{\nu}\Big[ C^n_{ij} -\frac{\mu_i\mu_j}{\nu}\Big]^{-1} \frac{\mu^c_j}{\nu} \Big) \notag\\&
=\Big(\nu-\mu_i^c (C^c_{ij})^{-1} \mu^c_j\Big)^{-1} \\&
=\Big(\nu-\mu_{\alpha\beta}^c (C^c_{\alpha\beta\gamma\delta})^{-1} \mu^c_{\gamma\delta} \Big)^{-1}\notag.&
\end{align}
As in Eq.~\eqref{eq43}, the second line followed from a Woodbury identity, and  the usual notation is used instead of the Voigt one in the last line. This covariance of the number  of point defects, will be connected to a compressibility-like expression $\kappa^c$ below.

\subsection{The isothermal compressibility of crystals}\label{isothermal}

In this section the general expression for the compressibility of a real crystal is derived from a thermodynamic consideration, details are given in Appendix \ref{appComp}. The situation described is one in which no pre-stress is applied to the crystal in equilibrium.

The definition of the isothermal compressibility of a fluid reads
\begin{equation}
\kappa =-\frac{1}{V} \frac{\partial V}{\partial p}\Big|_{N},
\end{equation}
where $p$ is the pressure. For a crystal, the question arises how this has to be generalized to describe the additional degrees of freedom. The infinitesimal change of the free energy of a crystal at constant temperature ($dT=0$) 
\begin{equation}\label{firstlawF}
dF = -pdV +\mu dN + h_{\alpha\beta}dU_{\alpha\beta},
\end{equation}
includes a term with a stress tensor $h_{\alpha\beta}$ at constant volume $V$ and particle number $N$ times an extensive strain tensor $U_{\alpha\beta}=V u_{\alpha\beta}$. The work done is $\delta W = \int h_{\alpha\beta}\delta u_{\alpha\beta} dV$ with the symmetrized linear strain tensor $u_{\alpha\beta}=\frac{1}{2}(\nabla_\alpha u_\beta + \nabla_\beta u_\alpha)$. The chemical potential is denoted by $\mu$, and the particle density will be denoted $n$.
While this 'first law of thermodynamics' for a crystal is familiar from textbooks \cite{Chaikin95}, the coupling of strain and density fluctuations complicates the interpretation of the stress tensor $h_{\alpha\beta}$, which calls for a discussion before addressing the compressibility. Taking a canonical $N$-particle system and straining it infinitesimally \cite{Squire69,Lutsko89} leads to the stress tensor $t_{\alpha\beta}=\frac1V \frac{\partial F}{\partial u_{\alpha\beta}}$ at fixed $T$ and $N$.  (It can be also obtained from averaging the Irving-Kirkwood microscopic stress tensor.) Because the volume $V$ varies in this procedure, the two stress tensors differ by a scalar term: \cite{Fleming76,Fuchs2012} $t_{\alpha\beta}=h_{\alpha\beta} - 
\left(p - h_{\gamma\delta}u_{\gamma\delta}\right) \delta_{\alpha\beta}$.
 The compressibility for a periodic crystal shall be understood as the derivative at constant $h$-stress tensor, because it then measures the change in particle density with chemical potential,
\begin{align}
\kappa &= -\frac{1}{V} \frac{\partial V}{\partial p}\Big|_{N,h_{\alpha\beta}}
 = \frac{1}{n^2_0}\frac{\partial n}{\partial \mu}\Big|_{h_{\alpha\beta}},\label{compdef}
\end{align}
where we used Maxwell and Gibbs-Duhem relations described in Appendix \ref{appComp}.  They lead to the Gibbs fundamental form of the free energy density $f=F/V$ which was already anticipated in Eqs.~\eqref{thermoderiv}, namely:
\begin{equation}\label{dfeq}
df = \mu dn + h_{\alpha\beta} du_{\alpha\beta}\;. \end{equation}
Also, the calculations for determining $\kappa$ have already been done. Equations.~\eqref{eq42} and \eqref{eq43} immediately give the isothermal compressibility as variance of the total density fluctuations:
\begin{align}\label{thermocomp}
\kappa &= \frac{1}{Vk_BT}\; \langle \frac{\delta n \delta n}{n^2_0}\rangle\Big|_{q=0}. 
\end{align}
\rem{$= \nu^{-1}+ \nu^{-1}\mu_{\alpha\beta}\Big[ C^n_{\alpha\beta\gamma\delta} - \mu_{\alpha\beta}\nu^{-1}\mu_{\gamma\delta}\Big]^{-1} \mu_{\gamma\delta} \nu^{-1}$}

\subsubsection{Including density}\label{sectIIIC1}

While the result for $\kappa$ in terms of the elastic coefficients (\textit{viz.}~Eqs.~\eqref{eq43} and \eqref{thermocomp}) is useful for explicit evaluations, and will be used in Sect.~\ref{cluster} below, a  relation connecting it to thermodynamic derivatives is desirable and would take the  form expected in the thermodynamic formalism. Using the relations \eqref{thermoderiv} in order to replace the elastic coefficients in Eq.~\eqref{eq43}, we find 
\rem{\begin{align*}
\kappa &= \Big(n_0^2\frac{\partial \mu}{\partial n}\Big|_{u_{\alpha\beta}}\Big)^{-1} +\frac{1}{n_0^2} \Big(\frac{\partial \mu}{\partial n}\Big|_{u_{\alpha\beta}}\Big)^{-1} \frac{\partial \mu}{\partial u_{\alpha\beta}}\Big|_{n}\notag\\
 &\quad\!\times\! \Big( \frac{\partial h_{\gamma\delta}}{\partial u_{\alpha\beta}}\Big|_n \!\!+\! \frac{\partial h_{\gamma\delta}}{\partial n}\Big|_{u_{\alpha\beta}}\!\frac{\partial n}{\partial u_{\alpha\beta}}\Big|_\mu  \Big)^{\!-1}\! \frac{\partial h_{\gamma\delta}}{\partial n}\Big|_{u_{\alpha\beta}}\!\!\Big( \frac{\partial \mu}{\partial n}\Big|_{u_{\alpha\beta}}\Big)^{\!-1}\!\! ,
\end{align*}}
\begin{align}
\kappa &= \frac{1}{n_0^2}\Big(\frac{\partial \mu}{\partial n}\Big|_{u_{\alpha\beta}}-\frac{\partial h_{\alpha\beta}}{\partial n}\Big|_{u_{\alpha\beta}}\Big(\frac{\partial h_{\gamma\delta}}{\partial u_{\alpha\beta}}\Big|_n\Big)^{-1} \frac{\partial h_{\gamma\delta}}{\partial n}\Big|_{u_{\alpha\beta}}\Big)^{-1}\;.
\label{kompressibilitaet}
\end{align}
This expression for the isothermal compressibility of a general crystal generalizes results obtained for high symmetry crystals \cite{Zippelius80}.  Hence, together with Eqs.~\eqref{thermocomp} and \eqref{eq43} and Sect.~\ref{sect:correlations}, we achieved our main goal to establish the general connection between the isothermal compressibility of non-ideal crystals and the correlation functions of the fields of elasticity theory. The connection is derived from microscopic DFT.
See Appendix \ref{appComp} for an alternative formulation of Eq.~\eqref{kompressibilitaet} derived within the thermodynamic formalism, and corresponding to the first line of Eq.~\eqref{eq43}. (Eq.~\eqref{kompressibilitaet} corresponds to the second line of Eq.~\eqref{eq43}.) \rem{It supplements the derivation in Ref.~[\onlinecite{Walz10}] (recalled in Eq.~\eqref{numulambda}), where the equivalence of the hydrodynamic equations was used.}
  
 If the coupling between strain and density fluctuations in the result for the isothermal compressibility in Eq.~\eqref{thermocomp} is neglected, the second term vanishes and the compressibility $\kappa$ is given by  $\kappa^{-1} = \nu=n_0^2 \frac{\partial \mu}{\partial n}\Big|_{u_{\alpha\beta}} $, which plays the role of the inverse bulk modulus at constant strain.   While in regular solids, the coupling between strain and density in the free energy, $\boldsymbol{\mu}=\frac{\partial^2 f}{\partial n \partial {\bf u}}$, cannot be neglected and this approximation fails,  see Sect.~\ref{cluster} for a  system where it holds well. In order to dissect the contributions to the compressibility in detail for more regular crystals,  transforming to defect density is required.

\subsubsection{Including defect density}

If one considers the set of independent variables with the defect density $ c$  instead of the coarse-grained density $n$ with Eq.~\eqref{defect} simplifying to 
\begin{equation}
d n = - n_0 d u_{\alpha\alpha} - d c \; ,\label{defectrelation} 
\end{equation}
the manipulations leading from Eq.~\eqref{thermocomp} to Eq.~\eqref{kompressibilitaet} have to be adapted. The compressibility is given now in terms of derivatives  at constant defect density \cite{Walz10}.  The stress tensor $\sigma_{\alpha\beta}$ (with  $\sigma_{\alpha\beta} =h_{\alpha\beta}-n_0 \mu  \delta_{\alpha\beta}$) and the chemical potential $\mu$ now are functions of the strain tensor and the defect density combining to the Gibbs  fundamental form of the free energy density \cite{Fleming76,Fuchs2012} $df= -\mu dc + \sigma_{\alpha\beta} d u_{\alpha\beta}$.
The relevant thermodynamic derivatives are  given by Eqs.~\eqref{thermoderiv2} (see Eq.~\eqref{FreeEnC} for the free energy density), which need to be used in order to replace the elastic coefficients in Eq.~\eqref{eq43}. This leads to: 
\begin{align}
\kappa &= -\left( n_0^2\frac{\partial \mu}{\partial c}\Big|_{u_{\alpha\beta}} \right)^{-1}\!\!\!\! +\!\left(\!\delta_{\alpha\beta}-\left(n_0 \frac{\partial \mu}{\partial c }\Big|_{u_{\alpha\beta}} \right)^{-1}\!\!\!\!\frac{\partial\mu}{\partial u_{\alpha\beta}}\Big|_{c} \right) \notag\\
 &\qquad \times \left(\frac{\partial\sigma_{\gamma\delta}}{\partial u_{\alpha\beta}}\Big|_c - \frac{\partial \sigma_{\gamma\delta}}{\partial c }\Big|_{u_{\alpha\beta}} \left(\frac{\partial \mu}{\partial c}\Big|_{u_{\alpha\beta}}\right)^{-1}\frac{\partial\mu}{\partial u_{\alpha\beta}}\Big|_{c} \right)^{-1} \notag\\ &\qquad \times \left(\! \delta_{\gamma\delta}+\frac{1}{n_0}\frac{\partial\sigma_{\gamma\delta}}{\partial c}\Big|_{u_{\alpha\beta}}\left(\frac{\partial\mu}{\partial c}\Big|_{u_{\alpha\beta}}\right)^{-1}  \right)\\ \notag
&=-\left(n_0^2\frac{\partial\mu}{\partial c}\Big|_{u_{\alpha\beta}}+\left(\!n_0^2\frac{\partial\mu}{\partial c}\Big|_{u_{\alpha\beta}}\delta_{\alpha\beta}+n_0\frac{\partial\sigma_{\alpha\beta}}{\partial c}\Big|_{u_{\alpha\beta}} \right)\right. \notag\\ &\qquad \times \left[\frac{\partial\sigma_{\alpha\beta}}{\partial u_{\gamma\delta}}\Big|_c-n_0^2\frac{\partial\mu}{\partial c}\Big|_{u_{\alpha\beta}}\delta_{\alpha\beta}\delta_{\gamma\delta}\right. \notag\\ &\qquad \left. -n_0 \frac{\partial\sigma_{\alpha\beta}}{\partial c}\Big|_{u_{\gamma\delta}}\delta_{\gamma\delta}-n_0 \frac{\partial\sigma_{\gamma\delta}}{\partial c}\Big|_{u_{\gamma\delta}}\delta_{\alpha\beta}\right]^{-1}\notag\\ &\qquad \times \left. \left(\!n_0^2\frac{\partial\mu}{\partial c}\Big|_{u_{\gamma\delta}}\delta_{\gamma\delta}+n_0\frac{\partial\sigma_{\gamma\delta}}{\partial c}\Big|_{u_{\alpha\beta}} \right)  \right)^{-1}.
\end{align}
An interesting limit is now the vanishing of the coupling between the defect density and the strain field, $\frac{\partial^2 f}{\partial c\partial u_{\alpha\beta}} =0$; see Eq.~\eqref{eq44b}. This yields two independent contributions to the compressibility which shall be denoted $\kappa^{0}$ in this approximation
\begin{align}\label{eq54}
\kappa^{0} &= \nu^{-1} + (C^c_{\alpha\beta\gamma\delta})^{-1} \delta_{\alpha\beta}\delta_{\gamma\delta} 
=\nu^{-1}+\sum_{i,j=1}^3(C^c_{ij})^{-1} .
\end{align}
The first contribution $\nu^{-1}$ is due to the fluctuations of the defect density, whereas the second one $(C^c_{\alpha\beta\gamma\delta})^{-1} \delta_{\alpha\beta}\delta_{\gamma\delta}$ is due to independent fluctuations of the strain tensor, which agrees with the known result for a perfect crystal without external strain\cite{Wallace70}.

\subsection{The isothermal defect density susceptibility}

Varying the chemical potential changes not only the average density  but also the defect density.
The derivative  of the defect density with respect to $\mu$ can be obtained analogously to Eq.~\eqref{thermocomp}, and  a  thermodynamic susceptibility akin to the compressibility can be defined 
\begin{align}\label{defectcomp}&
\kappa^{c} = \frac{-1}{n^2_0}\frac{\partial c}{\partial \mu}\Big|_{\sigma_{\alpha\beta}} = 
\langle \frac{\delta c \delta c}{Vk_BTn^2_0}\rangle\Big|_{q=0} 
\end{align}
The explicit result for $\kappa^{c}$ in terms of the elastic coefficients is given in Eq.~\eqref{kappac} and in terms of the derivatives from Eq.~\eqref{thermoderiv2} is given here:
\begin{align}\label{defectcomp2}
\kappa^c &= \frac{1}{n_0^2}\Big(-\frac{\partial \mu}{\partial c}\Big|_{u_{\alpha\beta}}-\frac{\partial \sigma_{\alpha\beta}}{\partial c}\Big|_{u_{\alpha\beta}}\Big(\frac{\partial \sigma_{\gamma\delta}}{\partial u_{\alpha\beta}}\Big|_c\Big)^{-1} \frac{\partial \sigma_{\gamma\delta}}{\partial c}\Big|_{u_{\alpha\beta}}\Big)^{-1}
\end{align}

Connecting the isothermal defect density susceptibility to  derivatives of the density appears useful in order to obtain it \textit{e.g.}~from computer simulations. Starting from the definition of $\kappa^c$ in  Eq.~\eqref{defectcomp}, the  Eq.~\eqref{defectrelation} leads to
\begin{align}
\frac{-1}{n^2_0}\frac{\partial c}{\partial \mu}\Big|_{\sigma_{\alpha\beta}} &= 
\frac{-1}{n^2_0}\frac{\partial c}{\partial n}\frac{\partial n}{\partial \mu}\Big|_{\sigma_{\alpha\beta}} +
\frac{-1}{n^2_0}\frac{\partial c}{\partial u_{\alpha\beta}}\frac{\partial u_{\alpha\beta}}{\partial \mu}\Big|_{\sigma_{\alpha\beta}}\nonumber\\
&=\frac{1}{n^2_0}\frac{\partial n}{\partial \mu}\Big|_{\sigma_{\alpha\beta}} +
\frac{1}{n_0}\frac{\partial u_{\alpha\alpha}}{\partial \mu}\Big|_{\sigma_{\alpha\beta}}
\end{align}
The first term on the right hand side is a thermodynamic susceptibility at constant $\sigma$-stress tensor, which bears similarity to the isothermal compressibility: 
\begin{align}
\kappa^{\sigma}=\frac{1}{n^2_0}\frac{\partial n}{\partial \mu}\Big|_{\sigma_{\alpha\beta}}
\end{align}
Yet, Appendix \ref{appElast} will show that this specific susceptibility vanishes in the limit of an ideal crystal, and thus does not play the role of a compressibility in solids. The second term can be reformulated using Eq.~\eqref{Matrix_c}, and the result can be rearranged to give: 
\begin{align}\label{eq60}
\kappa^{\sigma} &=\kappa^c\Big(1-\frac{\partial \sigma_{\gamma\delta}}{\partial c}\Big|_{u_{\alpha\beta}}\Big(\frac{\partial \sigma_{\alpha\alpha}}{\partial u_{\gamma\delta}}\Big|_c\Big)^{-1}\Big)\\
&=\kappa^c(1-\mu^c_{\gamma\delta}(C^c_{\gamma\delta\alpha\alpha})^{-1})
=\kappa^c(1-\sum_{j=1}^3\mu^c_i(C^c_{ij})^{-1}),\notag
\end{align}
where in the last equality the thermodynamic derivatives from Eq.~\eqref{thermoderiv2}  were used, and the result transferred in Voigt notation. The difference between $\kappa^c$ and $\kappa^\sigma$, which both are derivatives at constant ${\boldsymbol \sigma}$-stress tensor, vanishes in cases where strain and defect density fluctuations do not couple (viz.~$\boldsymbol \mu^c=0$). In the general case, density and (the negative of the) defect density vary differently with chemical potential at fixed ${\boldsymbol \sigma}$.

\section{Small wavevector limit of the structural functions}\label{sect:small}

So far we considered correlation functions and the isothermal compressibility of crystals. In this section we bridge the gap between the density correlation function and the compressibility, and point out the subtle difference between the two expressions. In the second part of this chapter the so called generalized structure factor is discussed.

In order to understand the connection to the compressibility, the $q$-dependence in the limit $q\to 0$ of the correlation function of the coarse-grained density \eqref{dichtedichte} needs to be discussed in detail
\begin{align}
\langle & \frac{\delta n^\ast\delta n}{Vk_BT n_0^2}\rangle = \nu^{-1}(\mathbf{q}) + \nu^{-1}(\mathbf{q})\mu_\alpha^\ast(\mathbf{q}) \notag\\
&\times \Big(\lambda_{\alpha\gamma}(\mathbf{q})-\mu_\alpha(\mathbf{q}) \nu^{-1}(\mathbf{q})\mu_\gamma^\ast(\mathbf{q}) \Big)^{-1}\mu_\gamma(\mathbf{q}) \nu^{-1}(\mathbf{q})\notag \\
&\stackrel{q\to 0}{=} \frac 1\nu +\! \frac{\mu_{\alpha\beta}q_\beta}{\nu} \left[(\lambda_{\alpha\gamma\epsilon\phi}-\frac{\mu_{\alpha\epsilon}\mu_{\gamma\phi}}{\nu})q_\epsilon q_\phi \right]^{-1} \frac{\mu_{\gamma\delta}q_\delta}{\nu}.\label{eq57}
\end{align}
Here we used the known small-wave vector expansions of the elastic coefficients, which were defined in Eqs.~\eqref{numulambda}. They follow from DFT relations expressing translational and rotational symmetry\cite{Walz10}. Noting that only the symmetrized combinations in $\alpha\leftrightarrow\gamma$ and $\epsilon\leftrightarrow\phi$ of the term in square brackets are relevant, and with the help of Eqs.~\eqref{thermoderiv} this expression can be further simplified to
\begin{align}\label{dyncomp}
\langle \frac{\delta n^\ast\delta n}{Vk_BT n_0^2}\rangle &\stackrel{q\to 0}{=} \frac 1\nu \! +\! \frac{\mu_{\alpha\beta}q_\beta}{\nu}\! \left[(C^n_{\alpha\epsilon\gamma\phi}\!-\!\frac{\mu_{\alpha\epsilon}\mu_{\gamma\phi}}{\nu})q_\epsilon q_\phi \right]^{-1}\! \frac{\mu_{\gamma\delta}q_\delta}{\nu}.
\end{align}

This expression would agree with the thermodynamic one \eqref{thermocomp}, if the factors of $q_\beta q_\delta$ canceled $q_\epsilon q_\phi$. 
That the limit $q\to 0$ is not that simple can be seen even for highly symmetric crystals. For a cubic crystal, the correlation function yields different results in the small $q$ limit \eqref{dyncomp} depending on the direction of $\bf q$ relative to the unit cell. And for the hypothetical model of an isotropic crystal, the  small $q$ limit \eqref{dyncomp} is direction independent, but differs from the thermodynamic value from \eqref{thermocomp}. The latter simplified case, allows to identify the origin of the discrepancy and will be studied in detail in the next section. 

\subsection{Perfect crystal embedded in a matrix}

To study the difference in more detail, it is, as a first simplification, more convenient to look at the simpler problem of a perfect crystal. In this section we also use the more familiar expressions of elasticity theory. The connection to the terms used so far is given in Appendix \ref{appElast}. For a perfect crystal  the correlations of the displacement field is given by the (inverse) of the dynamical matrix $D_{\alpha\beta}(\mathbf{q})$ (for particles with mass $m$)
\begin{align}\label{PerKorrU}
\langle \delta u_\alpha^\ast \delta u_\beta\rangle &= \frac{Vk_BT}{mn_0} \;  D_{\alpha\beta}^{-1}(\mathbf{q}).
\end{align}
The coarse-grained density fluctuation for a perfect crystal is $\delta n(\mathbf{q},t) = -i n_0 q_\alpha \delta u_\alpha(\mathbf{q},t)$ and the dynamical matrix is related with the elastic constants\cite{Chaikin95} via $D_{\alpha\gamma}(\mathbf{q})= C_{\alpha\beta\gamma\delta}q_\beta q_\delta$. Thus for the coarse-grained density correlation function we get
\begin{align}\label{PerKorrN}
\langle \frac{\delta n^\ast \delta n}{Vk_BT n_0^2}\rangle &= q_\alpha D^{-1}_{\alpha\beta}(\mathbf{q}) q_\beta / (mn_0 ) = q_\alpha (C_{\alpha\epsilon\beta\phi} q_\epsilon q_\phi)^{-1} q_\beta
\end{align}
which shows the same problem in the limit $q\to 0$ as arises in Eq.~\eqref{dyncomp}, when compared to the thermodynamic compressibility of an ideal crystal\cite{Wallace70} $\kappa^{\rm ic} = (C^{-1}_{\alpha\beta\gamma\delta}) \delta_{\alpha\beta} \delta_{\gamma\delta}$ (contraction of the inverse of the matrix of elastic constants). For an isotropic crystal the elastic tensor  simplifies to the two Lam\'{e} coefficients $C_{\alpha\beta\gamma\delta}=\lambda \delta_{\alpha\beta}\delta_{\gamma\delta}+\mu(\delta_{\alpha\gamma}\delta_{\beta\delta}+\delta_{\alpha\delta}\delta_{\beta\gamma})$. Thus, the compressibility is $(\kappa^{\rm ic})^{-1} = \lambda +\frac{2}{3}\mu$, whereas the correlation function yields $\lambda + 2\mu$ (which corresponds to the longitudinal speed of sound).

To show the origin of this difference we consider an isotropic (ideal) solid for which the so called fundamental solution of elasticity is known. Other symmetries with known solutions are hexagonal\cite{Kroener53} and pentagonal\cite{De87}. The corresponding problem in two dimensions can be found in [\onlinecite{Franzrahe10}]. We consider a three dimensional sphere with volume $V_B$ embedded in a spherical matrix $V$ of the same isotropic material. The radius $R_B$ of the embedded sphere is increased $R_B\to R_B + \Delta r$ and the surrounding matrix is compressed. To determine the displacement field and the elastic energy of such a deformation one has to solve the equation of elastostatic theory, which is the vanishing of the divergence of the stress tensor, or in terms of displacement field
\begin{align}\label{GrundglElastostatik}
\nabla_\beta C_{\alpha\beta\gamma\delta} \nabla_\gamma u_\delta = 0.
\end{align}
The solution is a sphere with increased volume $V_B+\Delta V_B$. The only non-vanishing displacement field is (homogeneous dilatation)
\begin{align}
\delta u_r &= \left\{ \begin{array}{c} \Delta r \frac{r}{R_B} \qquad\qquad\!  r< R_B\\ \Delta r \left(\frac{R_B}{r}\right)^2 \qquad r > R_B \end{array}\right.
\end{align}
This yields for the total elastic energy
\begin{align}
E = \frac{V_B}{2}\Big(\frac{\Delta V_B}{V_B}\Big)^2\Big[ (\lambda +\frac{2}{3}\mu) + \frac{4}{3}\mu \Big( 1-\frac{V_B}{V}\Big)\Big].
\end{align}
The first part is due to the stretched sphere and the second contribution is from the surrounding matrix. Thus, depending on the ratio $\frac{V_B}{V}$ the relevant combination of elastic constants changes from $\lambda + \frac{2}{3}\mu$ for ($\frac{V_B}{V}\to 1$) to $\lambda + 2\mu$ for ($\frac{V_B}{V} \to 0$). In the limit of vanishing shear modulus $\mu$ the difference vanishes. Thus, for a fluid it doesn't matter if one determines the volume fluctuations of a small sphere in surrounding fluid (of the same kind) or if one looks at the global fluctuations of the whole system.

It is worthwhile to note that the same ratio between these two combinations of Lam\'e coefficients appears in a related context. In Eshelby's study\cite{Eshelby57} of an inclusion in a matrix of elastic material, the so called constrained strain $u^C_{\alpha\beta}$ is given by the stress-free strain $u^T_{\alpha\beta}$
\begin{align}
u^C_{\alpha\alpha} &= \frac{\lambda + \frac{2}{3}\mu}{\lambda + 2\mu} u^T_{\alpha\alpha}.
\end{align}
This calculation has recently been extended to atomistic inclusions  \cite{Garikipati06}, which could be used to test approximations in the present DFT approach.
An analogous problem is a polar fluid in a dielectric medium\cite{Hansen96,Titulaer74,Nienhuis71}. There, the susceptibilities show a directional dependence due to the dipolar interaction, and a different combination of dielectric constants is relevant depending on the surrounding medium.

\subsection{Generalized Structure Factor}

There is a further aspect which differs the relation between the compressibility and the correlation of the density fluctuations of a fluid and a crystal. There is a difference if one looks at the elements of the generalized structure factor which contribute to the compressibility, i.e. whether those are different from $S_{\mathbf{g=0}}(\mathbf{q}\to 0)$. 

We recall that the generalized structure factor is defined by \cite{McCarley}
\begin{align}
S_\mathbf{g}(\mathbf{k}) &= \frac 1V \int d^d\!r_1\!\int d^d\!r_2 \ \langle \delta\rho(\mathbf{r_1})\delta\rho(\mathbf{r_2})\rangle\ e^{-i\mathbf{g\cdot R}} \ e^{-i\mathbf{k\cdot \Delta r}} \notag\\
&= \frac 1V \langle \delta\rho(\mathbf{g/2 +k}) \delta\rho(\mathbf{g/2 -k}) \rangle\;,\label{strukturfaktor}
\end{align}
(with $\mathbf{R}=(\mathbf{r_1}+\mathbf{r_2})/2$ and $\mathbf{\Delta r = r_1-r_2}$) and its $S_0(\tilde{\mathbf{g}}+\mathbf{q})$ element is measured in a scattering experiment\cite{Chaikin95,Ashcroft76}. 

In a liquid, where translational invariance dictates that only $S_0(k)$ is non-vanishing and isotropic, its connection\cite{Hansen96,Barrat03} to the compressibility is well known $S_0(q\to0)\to n_0^2 k_BT\kappa$.
To convince oneself that such a connection does not hold in a crystal, the definition of the coarse-grained density Eq.~(\ref{macrodensity}) can be  used to derive
\begin{align}
\left\langle \frac{\delta n^\ast \delta n}{n_0^2} \right\rangle \! &= \frac{1}{\mathcal{N}^2_0} \sum_{\mathbf{g,g^\prime}} n_{\mathbf{g^\prime}} \langle \delta\rho^\ast(\mathbf{g^\prime +q}) \delta \rho(\mathbf{g+q})\rangle n^\ast_\mathbf{g}\notag\\
&= \!\frac{(2\pi)^d}{\mathcal{N}^2_0}\!\! \sum_{\mathbf{g,g^\prime}}\!\! n_{\mathbf{g^\prime}}\!\! \sum_{\mathbf{\tilde{g}}}\! S_{\mathbf{\tilde{g}}}(\frac{\mathbf{\tilde{g}}}{2}\!-\mathbf{g\!-q}) \delta(\mathbf{g-g^\prime-\tilde{g}}) n_\mathbf{g}^\ast \notag\\
&= \frac{(2\pi)^d}{\mathcal{N}^2_0} \sum_{\mathbf{g,g^\prime}}  n_{\mathbf{g^\prime}} S_{\mathbf{g-g^\prime}}\left(-\frac{\mathbf{g+g^\prime}}{2}-\mathbf{q}\right) n^\ast_\mathbf{g},
\end{align}
where the left hand side becomes $\kappa$ for $\bf q$ to zero in the fluid case.
Clearly, every element of $\langle \delta\rho^\ast(\mathbf{g^\prime +q}) \delta \rho(\mathbf{g+q})\rangle$ is involved, not just the one with vanishing reciprocal lattice vector $\mathbf{g}=\mathbf{g^\prime} =0$. Also the correlation of coarse-grained density fluctuations is given by a combination of generalized structure factors $S_\mathbf{g-g^\prime}(-(\mathbf{g+g^\prime})/2 -\mathbf{q})$ in the limit $\mathbf{q}\to 0$ and not just by $S_{\mathbf{\hat{g}=0}}(\mathbf{q}\to 0)$ as for a fluid. Although the possibility that the right-hand side of the last equation is indeed the compressibility cannot be ruled out, it seems rather unlikely.

\section{An example: Cluster crystals}\label{cluster}

To test the theory presented in the preceding sections, single component crystals of Bravais symmetry formed by spherical particles  provide the closest cases. Large densities of local defects are desirable  since the strength of the theory is its ability to account for the coupling of strain and defects densities. Additionally, a good approximate DFT functional should be available.  Recently, cluster crystals made from soft particles were discovered which satisfy these criteria and are thus ideally suited for testing the theory.

\subsection{Model and approximate density functional theory}

We consider a system of spherically symmetric particles interacting via a purely repulsive, bounded pair potential. Following earlier studies\cite{Mladek06,Mladek07}, we use a generalized exponential model of exponent four (GEM-4),
\begin{equation}
\Phi(r)=\epsilon e^{-(\frac{r}{\sigma})^{4}}.\\
\end{equation}
The GEM-4-system shows several interesting properties. The finite upper bound of the potential allows cluster formation, i.e. the occupation of one lattice
site by several particles. The GEM-4-system crystallizes in the fcc and bcc phases, and at low temperatures it undergoes isostructural phase transitions between fcc phases
with integer occupation numbers per lattice site. At higher temperatures hopping of the particles between the lattice sites is possible and leads
to a continuous, average occupation number. For the average density distribution of the cluster crystal the following ansatz is chosen\cite{Mladek06}
\begin{equation}
\rho(\mathbf{r})=n_{c}\left(\frac{\alpha}{\pi}\right)^{\frac{3}{2}}\sum_{\mathbf{R}}e^{-\alpha(\mathbf{r}-\mathbf{R})^{2}} 
\end{equation}
with the occupation number $n_{c}$, the  inverse width of the (Gaussian) density distribution around a single lattice site $\alpha$ and the lattice
vectors $\mathbf{R}$. With this ansatz and an appropriate free energy functional one can get the parameters $n_{c}$, and $\alpha$ for given temperatures,
and average densities through minimization of the functional. 
Then, the equilibrium state can be found by a direct comparison of the free
energies of each phase. For the description of the phase-diagram of the GEM-4, Mladek and coworkers found that a liquid-like mean-field approximation  is appropriate \cite{Likos07} which leads to the simple expression for the direct correlation function $c(\mathbf{r_{1}},\mathbf{r_{2}})$
\begin{equation}
c(\mathbf{r_{1}},\mathbf{r_{2}})\equiv c(r)=-\beta\Phi(r)\;, \mbox{with}\quad
r=|\mathbf{r_2-r_1}|.\end{equation}
This results in the following free energy functional 
\begin{align}
&F[\rho]=F_{\text{id}}[\rho]+F_{\text{ex}}[\rho],\\ \nonumber
&F_{\text{id}}[\rho]=\frac{1}{\beta}\int d^{3}\mathbf{r}[\rho(\mathbf{r})\ln[\rho(\mathbf{r})\Lambda^{3}]-\rho(\mathbf{r})],\\ \nonumber
&F_{\text{ex}}[\rho]=\frac{1}{2}\int d^{3}\mathbf{r_{1}}\rho(\mathbf{r_{1}})\int d^{3}\mathbf{r_{2}}\Phi(\mathbf{r_{1}},\mathbf{r_{2}})\rho(\mathbf{r_{2}}). \\ \nonumber
\end{align}
Here $\Lambda$ denotes the thermal de Broglie wavelength. By subtracting the free energy of the fluid from the crystal one, the parameter $\Lambda$ can be eliminated
without changing the position of the minimum of the crystal free energy functional. 
Similarly, dividing by the average density $n_{0}$ does not change the free energy functional minimum, but leads to a convenient expression 
\begin{align}&
\tilde{f}\left(\frac{n_{c}}{n_{0}\sigma^{3}},\alpha\sigma^{2},\frac{k_{B}T}{n_{0}\sigma^{3}\epsilon}\right)=\frac{\Delta f}{n_{0}\sigma^{3}}&\nonumber\\ 
&=\frac{k_{B}T}{n_{0}\sigma^{3}\epsilon}\left(\ln\frac{n_{c}}{n_{0}\sigma^{3}}+\frac{3}{2}\ln{\{ \frac{\alpha\sigma^{2}}{e\pi}\}}\right) +\frac{1}{2}\sum_{\mathbf{g}\neq 0}e^{-\frac{g^{2}}{2\alpha}}\Phi_{\mathbf{g}}, &\label{df} 
\end{align}
with the Fourier transformed  potential $\Phi_{\mathbf{g}}$. As we alluded to earlier, the free energy functional (\ref{df}) is to be minimized with respect to $n_c$ and $\alpha$. The resulting, normalized free energy only depends on the single (dimensionless)
thermodynamic parameter $\frac{k_{B}T}{n_{0}\sigma^{3}\epsilon}$. Thus, the fluid-bcc and the fcc-bcc phase transitions lie on straight lines drawn from the 
origin of the $T-n_{0}$ phase diagram. It should be noted that the free energy functional (\ref{df}) is minimized by the ratio $\frac{n_{c}}{n_{0}\sigma^{3}}$ 
instead of  $n_{c}$ itself. Numerical minimization shows that $\frac{n_{c}}{n_{0}\sigma^{3}}$ varies only by about $\pm 3\%$ in the whole solid phase,
i.e. the system changes its density mainly due to changes in the occupation number and not due to changes in the lattice constant.\cite{Likos07}

\subsection{Compressibility and occupation number covariance}

\begin{figure}[htb]
\begingroup
  \makeatletter
  \providecommand\color[2][]{%
    \GenericError{(gnuplot) \space\space\space\@spaces}{%
      Package color not loaded in conjunction with
      terminal option `colourtext'%
    }{See the gnuplot documentation for explanation.%
    }{Either use 'blacktext' in gnuplot or load the package
      color.sty in LaTeX.}%
    \renewcommand\color[2][]{}%
  }%
  \providecommand\includegraphics[2][]{%
    \GenericError{(gnuplot) \space\space\space\@spaces}{%
      Package graphicx or graphics not loaded%
    }{See the gnuplot documentation for explanation.%
    }{The gnuplot epslatex terminal needs graphicx.sty or graphics.sty.}%
    \renewcommand\includegraphics[2][]{}%
  }%
  \providecommand\rotatebox[2]{#2}%
  \@ifundefined{ifGPcolor}{%
    \newif\ifGPcolor
    \GPcolortrue
  }{}%
  \@ifundefined{ifGPblacktext}{%
    \newif\ifGPblacktext
    \GPblacktexttrue
  }{}%
  \let\gplgaddtomacro\g@addto@macro
  \gdef\gplbacktext{}%
  \gdef\gplfronttext{}%
  \makeatother
  \ifGPblacktext
    \def\colorrgb#1{}%
    \def\colorgray#1{}%
  \else
    \ifGPcolor
      \def\colorrgb#1{\color[rgb]{#1}}%
      \def\colorgray#1{\color[gray]{#1}}%
      \expandafter\def\csname LTw\endcsname{\color{white}}%
      \expandafter\def\csname LTb\endcsname{\color{black}}%
      \expandafter\def\csname LTa\endcsname{\color{black}}%
      \expandafter\def\csname LT0\endcsname{\color[rgb]{1,0,0}}%
      \expandafter\def\csname LT1\endcsname{\color[rgb]{0,1,0}}%
      \expandafter\def\csname LT2\endcsname{\color[rgb]{0,0,1}}%
      \expandafter\def\csname LT3\endcsname{\color[rgb]{1,0,1}}%
      \expandafter\def\csname LT4\endcsname{\color[rgb]{0,1,1}}%
      \expandafter\def\csname LT5\endcsname{\color[rgb]{1,1,0}}%
      \expandafter\def\csname LT6\endcsname{\color[rgb]{0,0,0}}%
      \expandafter\def\csname LT7\endcsname{\color[rgb]{1,0.3,0}}%
      \expandafter\def\csname LT8\endcsname{\color[rgb]{0.5,0.5,0.5}}%
    \else
      \def\colorrgb#1{\color{black}}%
      \def\colorgray#1{\color[gray]{#1}}%
      \expandafter\def\csname LTw\endcsname{\color{white}}%
      \expandafter\def\csname LTb\endcsname{\color{black}}%
      \expandafter\def\csname LTa\endcsname{\color{black}}%
      \expandafter\def\csname LT0\endcsname{\color{black}}%
      \expandafter\def\csname LT1\endcsname{\color{black}}%
      \expandafter\def\csname LT2\endcsname{\color{black}}%
      \expandafter\def\csname LT3\endcsname{\color{black}}%
      \expandafter\def\csname LT4\endcsname{\color{black}}%
      \expandafter\def\csname LT5\endcsname{\color{black}}%
      \expandafter\def\csname LT6\endcsname{\color{black}}%
      \expandafter\def\csname LT7\endcsname{\color{black}}%
      \expandafter\def\csname LT8\endcsname{\color{black}}%
    \fi
  \fi
  \setlength{\unitlength}{0.0350bp}%
  \begin{picture}(7200.00,5040.00)%
    \gplgaddtomacro\gplbacktext{%
      \csname LTb\endcsname%
      \put(726,594){\makebox(0,0)[r]{\strut{} 0.2}}%
      \put(726,1117){\makebox(0,0)[r]{\strut{} 0.3}}%
      \put(726,1639){\makebox(0,0)[r]{\strut{} 0.4}}%
      \put(726,2162){\makebox(0,0)[r]{\strut{} 0.5}}%
      \put(726,2685){\makebox(0,0)[r]{\strut{} 0.6}}%
      \put(726,3207){\makebox(0,0)[r]{\strut{} 0.7}}%
      \put(726,3730){\makebox(0,0)[r]{\strut{} 0.8}}%
      \put(726,4252){\makebox(0,0)[r]{\strut{} 0.9}}%
      \put(726,4775){\makebox(0,0)[r]{\strut{} 1}}%
      \put(858,374){\makebox(0,0){\strut{} 0}}%
      \put(2344,374){\makebox(0,0){\strut{} 0.05}}%
      \put(3830,374){\makebox(0,0){\strut{} 0.1}}%
      \put(5315,374){\makebox(0,0){\strut{} 0.15}}%
      \put(6801,374){\makebox(0,0){\strut{} 0.2}}%
      \put(20,2684){\rotatebox{-270}{\makebox(0,0){\strut{}$\kappa n_0^2\epsilon\sigma^3$}}}%
      \put(3830,154){\makebox(0,0){\strut{}$k_B T/(\epsilon n_0 \sigma^3)$}}%
      \put(3532,1587){\makebox(0,0){\strut{}$\kappa$}}%
      \put(3532,2580){\makebox(0,0){\strut{}$\kappa^0$}}%
      \put(3532,3625){\makebox(0,0){\strut{}$\kappa^c$}}%
      \put(4324,2946){\makebox(0,0)[l]{\strut{}fcc}}%
      \put(5215,2946){\makebox(0,0)[l]{\strut{}bcc}}%
      \put(6157,2946){\makebox(0,0)[l]{\strut{}fluid}}%
    }%
    \gplgaddtomacro\gplfronttext{}
    \gplbacktext
    \put(0,0){\includegraphics[scale=0.7]{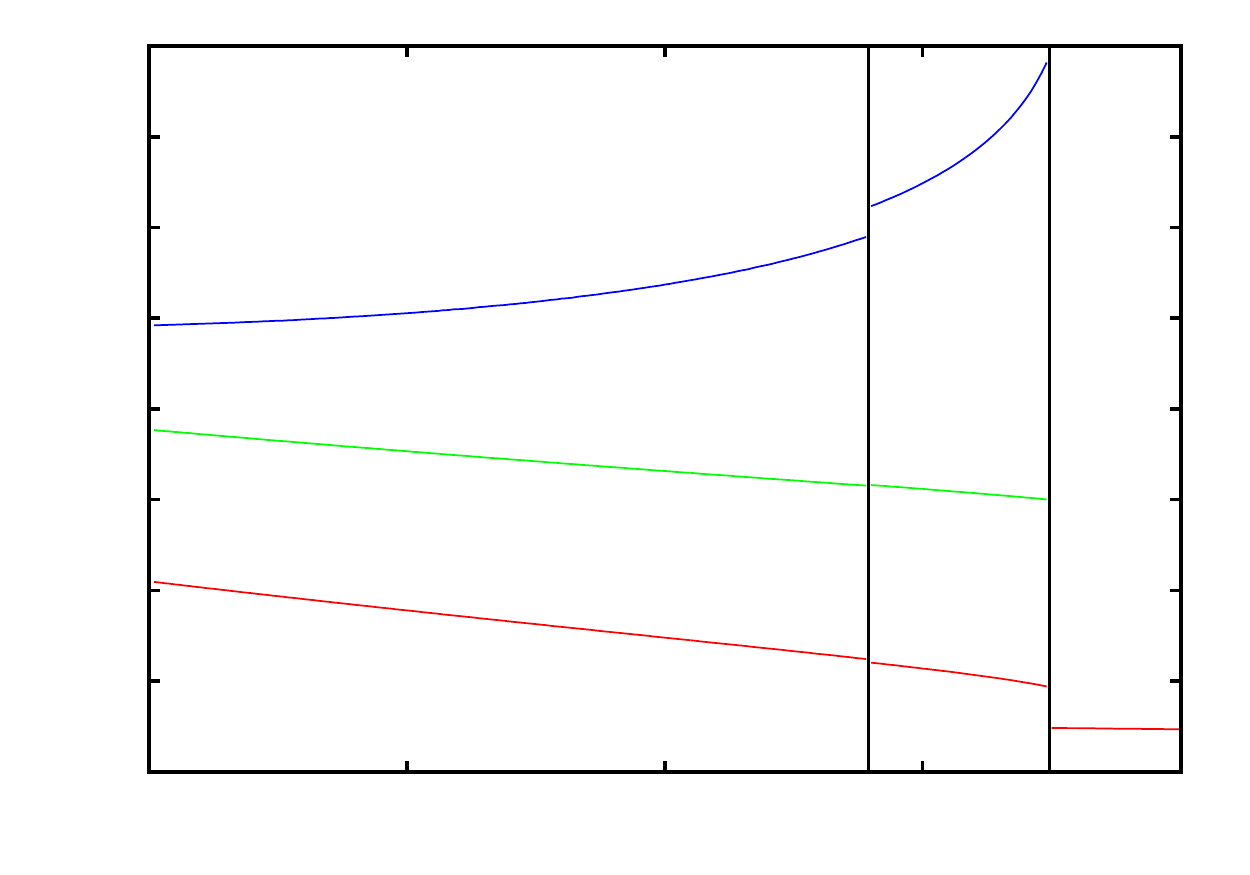}}%
    \gplfronttext
  \end{picture}%
\endgroup

\caption{Compressibilities of the GEM-4 system in units of $[n_{0}^{2}\epsilon\sigma^{3}]^{-1}$ versus the reduced thermodynamic variable $\frac{k_{B}T\sigma^{3}}{\epsilon n_{0}}$.  While $\kappa$ is taken at fixed stress  $h_{\alpha\beta}$, $\kappa^c$ is taken at fixed stress  $\sigma_{\alpha\beta}$, and $\kappa^0$ is the approximation neglecting the strain-density coupling introduced in Eq.~\eqref{eq54}. The approximation $\kappa\approx1/\nu$ to neglect the strain-defect density coupling holds within the line thickness; see Fig.\ref{fig4n}.}  \label{fig1}
\end{figure}

After minimizing the free energy and obtaining the average density profile, the elastic coefficients from Eq.~\ref{numulambda}, which are relevant for the compressibilities, can be calculated by straightforward integrations in the reciprocal space.  The thermodynamic derivatives then follow from the relations in Sect.~\ref{isothermal}.
Figure~\ref{fig1} shows three compressibility like quantities  in all stable phases obtained from the mean-field DFT functional \eqref{df}.  The compressibility $\kappa$ is taken at fixed stress  $h_{\alpha\beta}$ and describes the density change with chemical potential $\mu$. The susceptibility $\kappa^c$ is taken at fixed stress  $\sigma_{\alpha\beta}$ and captures the defect density change with $\mu$. The quantities $\kappa^0$ and $1/\nu$ are approximations neglecting the strain-density and strain-defect density coupling, respectively. In reduced units, the thermodynamic derivatives change little throughout the complete stable fcc phase. In the bcc crystal, defect fluctuations grow appreciably with increasing temperature. The fluid is less compressible than the solids, as follows from the familiar expression of the isothermal compressibility\cite{Hansen96}, $\kappa^{\rm fluid}= 1/\nu= (\partial n/\partial \mu)/n_0^2$.  The neglect of the coupling between strain and defect density qualitatively fails in the crystal phases. The full compressibility $\kappa$ differs strongly from the approximation $\kappa^0$, where both fields are assumed uncorrelated. Thus, widely made approximation\cite{Wallace70} which identifies  $\kappa$ and  $\kappa^0$
fails for cluster crystals. The very close agreement between $\kappa$ and $1/\nu$, on the other hand, indicates that the coupling between strain and density fluctuations is negligible, \textit{i.e.} $\mu_{\alpha\beta}\approx 0$; see Sect.~\ref{sectIIIC1}. Cluster crystals predominantly accommodate density changes by increasing the occupation numbers while keeping the lattice constants fixed \cite{Mladek06}. With the approximation $\mu_{\alpha\beta}\approx 0$, the coefficient $\mu^c_{\alpha\beta}$ becomes $\mu^c_{\alpha\beta}\approx\nu\delta_{\alpha\beta}$ and the formulas for $\kappa^c$ and $\kappa^{\sigma}$ simplify to
$\kappa^c \approx \nu^{-1}+\delta_{\alpha\beta}(C^n_{\alpha\beta\gamma\delta})^{-1}\delta_{\gamma\delta}$ and 
$\kappa^{\sigma}\approx\kappa\approx \nu^{-1}$.
Density changes with chemical potential similarly at fixed $\bf h$ and $\boldsymbol{\sigma}$ stress tensors. 
This is in strong contrast to the ideal crystal where $\kappa^{\sigma}$ equals $\kappa^c$ and both vanish.
\begin{figure}[htb]
\begingroup
  \makeatletter
  \providecommand\color[2][]{%
    \GenericError{(gnuplot) \space\space\space\@spaces}{%
      Package color not loaded in conjunction with
      terminal option `colourtext'%
    }{See the gnuplot documentation for explanation.%
    }{Either use 'blacktext' in gnuplot or load the package
      color.sty in LaTeX.}%
    \renewcommand\color[2][]{}%
  }%
  \providecommand\includegraphics[2][]{%
    \GenericError{(gnuplot) \space\space\space\@spaces}{%
      Package graphicx or graphics not loaded%
    }{See the gnuplot documentation for explanation.%
    }{The gnuplot epslatex terminal needs graphicx.sty or graphics.sty.}%
    \renewcommand\includegraphics[2][]{}%
  }%
  \providecommand\rotatebox[2]{#2}%
  \@ifundefined{ifGPcolor}{%
    \newif\ifGPcolor
    \GPcolortrue
  }{}%
  \@ifundefined{ifGPblacktext}{%
    \newif\ifGPblacktext
    \GPblacktexttrue
  }{}%
  \let\gplgaddtomacro\g@addto@macro
  \gdef\gplbacktext{}%
  \gdef\gplfronttext{}%
  \makeatother
  \ifGPblacktext
    \def\colorrgb#1{}%
    \def\colorgray#1{}%
  \else
    \ifGPcolor
      \def\colorrgb#1{\color[rgb]{#1}}%
      \def\colorgray#1{\color[gray]{#1}}%
      \expandafter\def\csname LTw\endcsname{\color{white}}%
      \expandafter\def\csname LTb\endcsname{\color{black}}%
      \expandafter\def\csname LTa\endcsname{\color{black}}%
      \expandafter\def\csname LT0\endcsname{\color[rgb]{1,0,0}}%
      \expandafter\def\csname LT1\endcsname{\color[rgb]{0,1,0}}%
      \expandafter\def\csname LT2\endcsname{\color[rgb]{0,0,1}}%
      \expandafter\def\csname LT3\endcsname{\color[rgb]{1,0,1}}%
      \expandafter\def\csname LT4\endcsname{\color[rgb]{0,1,1}}%
      \expandafter\def\csname LT5\endcsname{\color[rgb]{1,1,0}}%
      \expandafter\def\csname LT6\endcsname{\color[rgb]{0,0,0}}%
      \expandafter\def\csname LT7\endcsname{\color[rgb]{1,0.3,0}}%
      \expandafter\def\csname LT8\endcsname{\color[rgb]{0.5,0.5,0.5}}%
    \else
      \def\colorrgb#1{\color{black}}%
      \def\colorgray#1{\color[gray]{#1}}%
      \expandafter\def\csname LTw\endcsname{\color{white}}%
      \expandafter\def\csname LTb\endcsname{\color{black}}%
      \expandafter\def\csname LTa\endcsname{\color{black}}%
      \expandafter\def\csname LT0\endcsname{\color{black}}%
      \expandafter\def\csname LT1\endcsname{\color{black}}%
      \expandafter\def\csname LT2\endcsname{\color{black}}%
      \expandafter\def\csname LT3\endcsname{\color{black}}%
      \expandafter\def\csname LT4\endcsname{\color{black}}%
      \expandafter\def\csname LT5\endcsname{\color{black}}%
      \expandafter\def\csname LT6\endcsname{\color{black}}%
      \expandafter\def\csname LT7\endcsname{\color{black}}%
      \expandafter\def\csname LT8\endcsname{\color{black}}%
    \fi
  \fi
  \setlength{\unitlength}{0.03500bp}%
  \begin{picture}(7200.00,5040.00)%
    \gplgaddtomacro\gplbacktext{%
      \csname LTb\endcsname%
      \put(946,704){\makebox(0,0)[r]{\strut{} 0}}%
      \put(946,1518){\makebox(0,0)[r]{\strut{} 50}}%
      \put(946,2332){\makebox(0,0)[r]{\strut{} 100}}%
      \put(946,3147){\makebox(0,0)[r]{\strut{} 150}}%
      \put(946,3961){\makebox(0,0)[r]{\strut{} 200}}%
      \put(946,4775){\makebox(0,0)[r]{\strut{} 250}}%
      \put(1078,484){\makebox(0,0){\strut{} 3}}%
      \put(2032,484){\makebox(0,0){\strut{} 4}}%
      \put(2986,484){\makebox(0,0){\strut{} 5}}%
      \put(3941,484){\makebox(0,0){\strut{} 6}}%
      \put(4895,484){\makebox(0,0){\strut{} 7}}%
      \put(5849,484){\makebox(0,0){\strut{} 8}}%
      \put(6803,484){\makebox(0,0){\strut{} 9}}%
      \put(176,2739){\rotatebox{-270}{\makebox(0,0){\strut{}$B\sigma^3/\epsilon$}}}%
      \put(3940,154){\makebox(0,0){\strut{}$n_0\sigma^3$}}%
    }%
    \gplgaddtomacro\gplfronttext{%
      \csname LTb\endcsname%
      \put(3516,4602){\makebox(0,0)[r]{\strut{}$k_BT/\epsilon=0.5$}}%
      \csname LTb\endcsname%
      \put(3516,4382){\makebox(0,0)[r]{\strut{}$k_BT/\epsilon=0.8$}}%
      \csname LTb\endcsname%
      \put(3516,4162){\makebox(0,0)[r]{\strut{}$k_BT/\epsilon=1.1$}}%
      \csname LTb\endcsname%
      \put(3516,3942){\makebox(0,0)[r]{\strut{}MC}}%
    }%
    \gplbacktext
    \put(0,0){\includegraphics[scale=0.7]{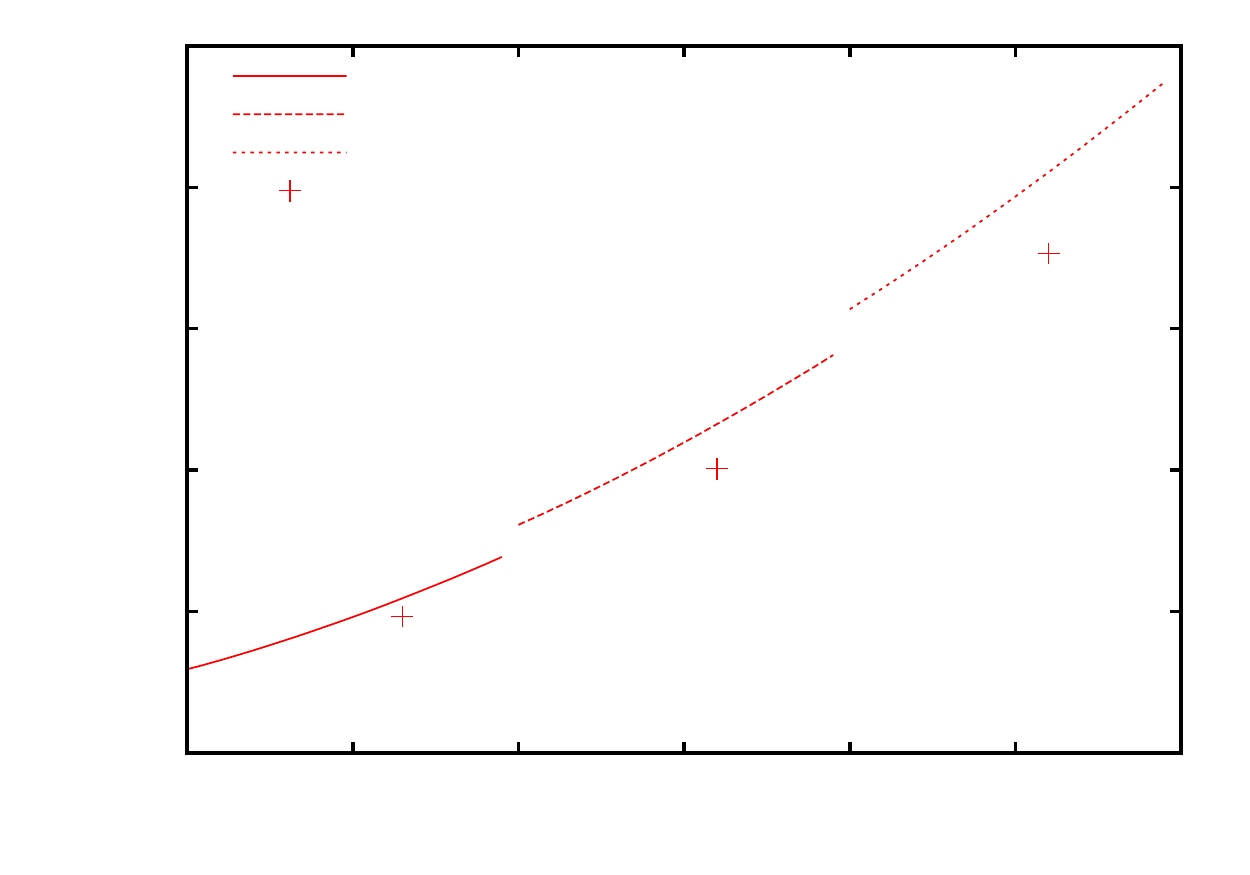}}%
    \gplfronttext
  \end{picture}%
\endgroup

\caption{The bulk modulus $B=\frac{1}{\kappa}$ in dimensionless units for three different temperatures versus $n_{0}\sigma^{3}$. The three points are MC simulation results\cite{Mladek07}.} \label{fig2}
\end{figure}

For a comparison with Monte Carlo (MC) simulations the  
compressibility $\kappa$ from Eq.~\eqref{thermocomp} is identified as inverse bulk modulus $B$ obtained in Ref.~[\onlinecite{Mladek07}]. Figure~\ref{fig2} shows this  bulk modulus for three
temperatures versus the average density. The deviation of the theoretical predictions from the simulation data by about $15\%$ is in the same range as the deviation of the calculated fcc-bcc transitions
from the simulated\cite{Mladek07} ones; this error is roughly $10\%$.

\begin{figure}[htb]
 \centering
\begingroup
  \makeatletter
  \providecommand\color[2][]{%
    \GenericError{(gnuplot) \space\space\space\@spaces}{%
      Package color not loaded in conjunction with
      terminal option `colourtext'%
    }{See the gnuplot documentation for explanation.%
    }{Either use 'blacktext' in gnuplot or load the package
      color.sty in LaTeX.}%
    \renewcommand\color[2][]{}%
  }%
  \providecommand\includegraphics[2][]{%
    \GenericError{(gnuplot) \space\space\space\@spaces}{%
      Package graphicx or graphics not loaded%
    }{See the gnuplot documentation for explanation.%
    }{The gnuplot epslatex terminal needs graphicx.sty or graphics.sty.}%
    \renewcommand\includegraphics[2][]{}%
  }%
  \providecommand\rotatebox[2]{#2}%
  \@ifundefined{ifGPcolor}{%
    \newif\ifGPcolor
    \GPcolortrue
  }{}%
  \@ifundefined{ifGPblacktext}{%
    \newif\ifGPblacktext
    \GPblacktexttrue
  }{}%
  \let\gplgaddtomacro\g@addto@macro
  \gdef\gplbacktext{}%
  \gdef\gplfronttext{}%
  \makeatother
  \ifGPblacktext
    \def\colorrgb#1{}%
    \def\colorgray#1{}%
  \else
    \ifGPcolor
      \def\colorrgb#1{\color[rgb]{#1}}%
      \def\colorgray#1{\color[gray]{#1}}%
      \expandafter\def\csname LTw\endcsname{\color{white}}%
      \expandafter\def\csname LTb\endcsname{\color{black}}%
      \expandafter\def\csname LTa\endcsname{\color{black}}%
      \expandafter\def\csname LT0\endcsname{\color[rgb]{1,0,0}}%
      \expandafter\def\csname LT1\endcsname{\color[rgb]{0,1,0}}%
      \expandafter\def\csname LT2\endcsname{\color[rgb]{0,0,1}}%
      \expandafter\def\csname LT3\endcsname{\color[rgb]{1,0,1}}%
      \expandafter\def\csname LT4\endcsname{\color[rgb]{0,1,1}}%
      \expandafter\def\csname LT5\endcsname{\color[rgb]{1,1,0}}%
      \expandafter\def\csname LT6\endcsname{\color[rgb]{0,0,0}}%
      \expandafter\def\csname LT7\endcsname{\color[rgb]{1,0.3,0}}%
      \expandafter\def\csname LT8\endcsname{\color[rgb]{0.5,0.5,0.5}}%
    \else
      \def\colorrgb#1{\color{black}}%
      \def\colorgray#1{\color[gray]{#1}}%
      \expandafter\def\csname LTw\endcsname{\color{white}}%
      \expandafter\def\csname LTb\endcsname{\color{black}}%
      \expandafter\def\csname LTa\endcsname{\color{black}}%
      \expandafter\def\csname LT0\endcsname{\color{black}}%
      \expandafter\def\csname LT1\endcsname{\color{black}}%
      \expandafter\def\csname LT2\endcsname{\color{black}}%
      \expandafter\def\csname LT3\endcsname{\color{black}}%
      \expandafter\def\csname LT4\endcsname{\color{black}}%
      \expandafter\def\csname LT5\endcsname{\color{black}}%
      \expandafter\def\csname LT6\endcsname{\color{black}}%
      \expandafter\def\csname LT7\endcsname{\color{black}}%
      \expandafter\def\csname LT8\endcsname{\color{black}}%
    \fi
  \fi
  \setlength{\unitlength}{0.03500bp}%
  \begin{picture}(7200.00,5040.00)%
    \gplgaddtomacro\gplbacktext{%
      \csname LTb\endcsname%
      \put(946,704){\makebox(0,0)[r]{\strut{}0}}%
      \put(946,1213){\makebox(0,0)[r]{\strut{}0.05}}%
      \put(946,1722){\makebox(0,0)[r]{\strut{}0.1}}%
      \put(946,2231){\makebox(0,0)[r]{\strut{}0.15}}%
      \put(946,2740){\makebox(0,0)[r]{\strut{}0.2}}%
      \put(946,3248){\makebox(0,0)[r]{\strut{}0.25}}%
      \put(946,3757){\makebox(0,0)[r]{\strut{}0.3}}%
      \put(946,4266){\makebox(0,0)[r]{\strut{}0.35}}%
      \put(946,4775){\makebox(0,0)[r]{\strut{}0.4}}%
      \put(1078,484){\makebox(0,0){\strut{} 8}}%
      \put(1880,484){\makebox(0,0){\strut{} 10}}%
      \put(2682,484){\makebox(0,0){\strut{} 12}}%
      \put(3484,484){\makebox(0,0){\strut{} 14}}%
      \put(4287,484){\makebox(0,0){\strut{} 16}}%
      \put(5089,484){\makebox(0,0){\strut{} 18}}%
      \put(5891,484){\makebox(0,0){\strut{} 20}}%
      \put(6693,484){\makebox(0,0){\strut{} 22}}%
      \put(176,2739){\rotatebox{-270}{\makebox(0,0){\strut{}$p(n)$}}}%
      \put(6912,2739){\rotatebox{-270}{\makebox(0,0){\strut{}}}}%
      \put(3885,154){\makebox(0,0){\strut{}occupancy $n$}}%
      \put(3885,4665){\makebox(0,0){\strut{}}}%
      \put(3885,4664){\makebox(0,0){\strut{}}}%
      \put(286,110){\makebox(0,0)[l]{\strut{}}}%
    }%
    \gplgaddtomacro\gplfronttext{%
      \csname LTb\endcsname%
      \put(2128,4492){\makebox(0,0)[l]{\strut{}bcc, $n_0\sigma^3=6.5$}}%
      \csname LTb\endcsname%
      \put(2128,4272){\makebox(0,0)[l]{\strut{}bcc, $n_0\sigma^3=7.5$}}%
      \csname LTb\endcsname%
      \put(2128,4052){\makebox(0,0)[l]{\strut{}fcc, $n_0\sigma^3=8.5$}}%
      \csname LTb\endcsname%
      \put(2128,3832){\makebox(0,0)[l]{\strut{}fcc, $n_0\sigma^3=9$}}%
    }%
    \gplbacktext
    \put(0,0){\includegraphics[scale=0.7]{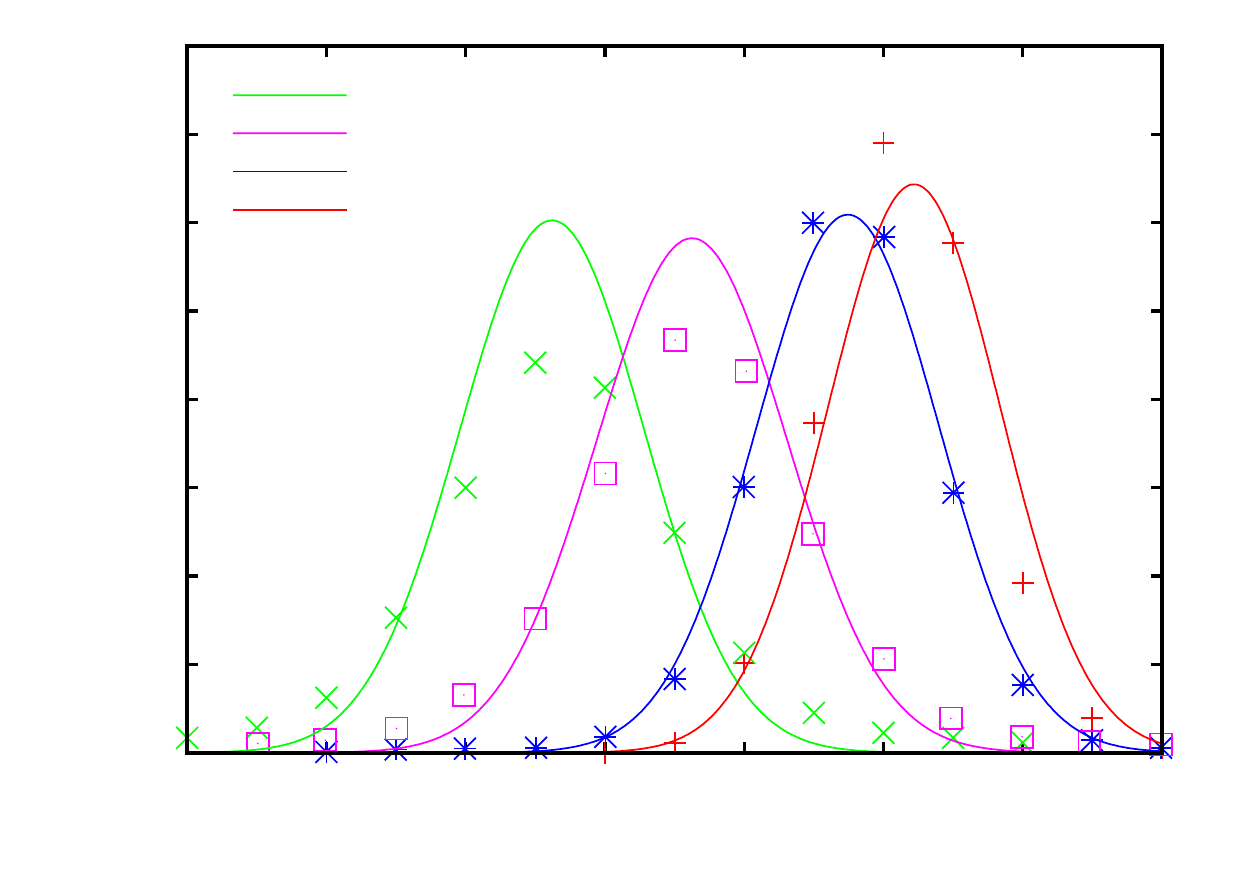}}%
    \gplfronttext
  \end{picture}%
\endgroup

 \caption{Probability distribution functions for the occupation numbers in GEM-4 cluster crystals of fcc and bcc structure from MC simulations\cite{Mladek06}. Gaussian distributions with the variances calculated from Eq.~\eqref{Deltanc2} and the mean value $n_c$ obtained through minimization of Eq.~\eqref{df} (lines) are compared with the MC data (symbols).\rem{The MC parameters are $k_{B}T/\epsilon=1$ with $n_{0}\sigma^{3}=9$, while the Gaussian has  (average) occupation number $n_{c}=18.4365$ and occupation number fluctuation $\sqrt{<\Delta n_{c}^{2}>}=1.2384$
as variance.} Complete parameters are given in  table I.}
 \label{fig3}
\end{figure}

The cluster crystal is an interesting model in the context of  defect density fluctuations. The  role of the defect density is taken by the  occupation number  $n_c$  which obviously is
an averaged number; it takes  real values, while a single lattice site can only be occupied with an integer number of particles. There has to be a distribution
in occupation numbers with the mean value $n_c$ and standard deviation $\sqrt{<\Delta n_{c}^{2}>}$ which should be connected with
$\langle \delta c \delta c\rangle$.  $\Delta n_c$ is the 
occupation number fluctuation for each lattice site, so the density $\delta c(\mathbf{r})$ has to be integrated over one primitive cell to become equivalent.
To simplify, we assume that the correlation in occupation number/defect density fluctuation vanishes after the first
Wigner-Seitz-cell, i.e. the occupation number fluctuation of each lattice site is independent.  
With ${N/n_{c}}$  the number of lattice sites
\begin{align}
V \int d^3r \langle \delta c(\mathbf{r}) \delta c(\mathbf{0})\rangle 
=\frac{N}{n_c}\langle\Delta n_c^2\rangle
\end{align}
This can be rewritten using  the compressibility $\kappa^c$ from Eq.~\ref{defectcomp}:
\begin{align} \langle\Delta n_c^2\rangle&=\kappa^c n^2_0 k_BT\left(\frac{n_c}{n_0}\right)\label{Deltanc2}
\end{align}
Assuming a Gaussian distribution, there is a good match of the results for the fcc lattice with MC simulations\cite{Mladek06}, as seen in  Fig.~\ref{fig3}. Table I collects the values of the averages obtained from the mean-field functional \eqref{df} and the variances obtained through Eq.~\eqref{Deltanc2}.
Also the percentage deviations from the parameters obtained from the Gaussian fits to the MC data are shown. The averages agree within 1\% for both lattices and the variances agree for the fcc lattice within 10\%, which is the same magnitude as for the Bulk modulus. For the bcc lattice bigger differences between the theoretical and the simulated \cite{Mladek06} occupation number distributions are observed for reasons unclear at present. The variances of defect fluctuations in bcc and fcc crystals are more different in the simulations than predicted theoretically.
 
\begin{table}[htb]
 \begin{tabular}{|c|c|c|c|c|c|c|c|c|}
\hline
\multicolumn{3}{|c}{}&\multicolumn{3}{|c}{$\sqrt{\langle\Delta n^2_c}\rangle$}&\multicolumn{3}{|c|}{$n_c$}\\
\hline
 \multicolumn{1}{|r|}{}&$k_BT/\epsilon$ & $n_0\sigma^3$& MC&  theory& $\Delta$[\%]&MC& theory& $\Delta$[\%]\\
\hline
bcc&1 & 6.5 &1.76&1.32&33.3&13.34&13.24&0.76\\
&1.1&7.5&1.66&1.37&21.2&15.31&15.25&0.39\\ 
\hline
fcc&1.1 & 8.5&1.23&1.31&6.5&17.48&17.49&0.06\\
&1 & 9 & 1.12&1.24&10.7&18.25&18.44&1.04\\	
\hline
 \end{tabular}
\caption{Variances and averages of the occupation numbers in cluster crystals with fcc and bcc  structure at selected state points. The MC results are obtained from Gaussian fits to Monte Carlo simulation data \cite{Mladek06}; the complete distributions are compared in Fig.~\ref{fig3}.
 The theoretical results for the averages follow from the mean-field DFT functional \eqref{df} and for the variances from \eqref{Deltanc2}. }
 \end{table}

\subsection{Dispersion relations and macroscopic density correlation function}

\begin{figure}[htb]
\begingroup
  \makeatletter
  \providecommand\color[2][]{%
    \GenericError{(gnuplot) \space\space\space\@spaces}{%
      Package color not loaded in conjunction with
      terminal option `colourtext'%
    }{See the gnuplot documentation for explanation.%
    }{Either use 'blacktext' in gnuplot or load the package
      color.sty in LaTeX.}%
    \renewcommand\color[2][]{}%
  }%
  \providecommand\includegraphics[2][]{%
    \GenericError{(gnuplot) \space\space\space\@spaces}{%
      Package graphicx or graphics not loaded%
    }{See the gnuplot documentation for explanation.%
    }{The gnuplot epslatex terminal needs graphicx.sty or graphics.sty.}%
    \renewcommand\includegraphics[2][]{}%
  }%
  \providecommand\rotatebox[2]{#2}%
  \@ifundefined{ifGPcolor}{%
    \newif\ifGPcolor
    \GPcolortrue
  }{}%
  \@ifundefined{ifGPblacktext}{%
    \newif\ifGPblacktext
    \GPblacktexttrue
  }{}%
  \let\gplgaddtomacro\g@addto@macro
  \gdef\gplbacktext{}%
  \gdef\gplfronttext{}%
  \makeatother
  \ifGPblacktext
    \def\colorrgb#1{}%
    \def\colorgray#1{}%
  \else
    \ifGPcolor
      \def\colorrgb#1{\color[rgb]{#1}}%
      \def\colorgray#1{\color[gray]{#1}}%
      \expandafter\def\csname LTw\endcsname{\color{white}}%
      \expandafter\def\csname LTb\endcsname{\color{black}}%
      \expandafter\def\csname LTa\endcsname{\color{black}}%
      \expandafter\def\csname LT0\endcsname{\color[rgb]{1,0,0}}%
      \expandafter\def\csname LT1\endcsname{\color[rgb]{0,1,0}}%
      \expandafter\def\csname LT2\endcsname{\color[rgb]{0,0,1}}%
      \expandafter\def\csname LT3\endcsname{\color[rgb]{1,0,1}}%
      \expandafter\def\csname LT4\endcsname{\color[rgb]{0,1,1}}%
      \expandafter\def\csname LT5\endcsname{\color[rgb]{1,1,0}}%
      \expandafter\def\csname LT6\endcsname{\color[rgb]{0,0,0}}%
      \expandafter\def\csname LT7\endcsname{\color[rgb]{1,0.3,0}}%
      \expandafter\def\csname LT8\endcsname{\color[rgb]{0.5,0.5,0.5}}%
    \else
      \def\colorrgb#1{\color{black}}%
      \def\colorgray#1{\color[gray]{#1}}%
      \expandafter\def\csname LTw\endcsname{\color{white}}%
      \expandafter\def\csname LTb\endcsname{\color{black}}%
      \expandafter\def\csname LTa\endcsname{\color{black}}%
      \expandafter\def\csname LT0\endcsname{\color{black}}%
      \expandafter\def\csname LT1\endcsname{\color{black}}%
      \expandafter\def\csname LT2\endcsname{\color{black}}%
      \expandafter\def\csname LT3\endcsname{\color{black}}%
      \expandafter\def\csname LT4\endcsname{\color{black}}%
      \expandafter\def\csname LT5\endcsname{\color{black}}%
      \expandafter\def\csname LT6\endcsname{\color{black}}%
      \expandafter\def\csname LT7\endcsname{\color{black}}%
      \expandafter\def\csname LT8\endcsname{\color{black}}%
    \fi
  \fi
  \setlength{\unitlength}{0.0400bp}%
  \begin{picture}(7200.00,5040.00)%
    \gplgaddtomacro\gplbacktext{%
      \csname LTb\endcsname%
      \put(594,440){\makebox(0,0)[r]{\strut{} 0}}%
      \put(594,1018){\makebox(0,0)[r]{\strut{} 2}}%
      \put(594,1596){\makebox(0,0)[r]{\strut{} 4}}%
      \put(594,2174){\makebox(0,0)[r]{\strut{} 6}}%
      \put(594,2752){\makebox(0,0)[r]{\strut{} 8}}%
      \put(594,3330){\makebox(0,0)[r]{\strut{} 10}}%
      \put(594,3908){\makebox(0,0)[r]{\strut{} 12}}%
      \put(594,4486){\makebox(0,0)[r]{\strut{} 14}}%
      \put(726,220){\makebox(0,0){\strut{}W}}%
      \put(1975,220){\makebox(0,0){\strut{}$\Gamma$}}%
      \put(3093,220){\makebox(0,0){\strut{}X}}%
      \put(220,2607){\rotatebox{-270}{\makebox(0,0){\strut{}$\omega\sigma\sqrt{m/\epsilon}$}}}%
      \put(3312,2607){\rotatebox{-270}{\makebox(0,0){\strut{}}}}%
      \put(1909,-66){\makebox(0,0){\strut{}}}%
      \put(1909,4665){\makebox(0,0){\strut{}}}%
      \put(1909,4664){\makebox(0,0){\strut{}}}%
      \put(66,110){\makebox(0,0)[l]{\strut{}}}%
    }%
    \gplgaddtomacro\gplfronttext{%
      \csname LTb\endcsname%
      \put(2406,4602){\makebox(0,0)[r]{\strut{}longitudinal}}%
      \csname LTb\endcsname%
      \put(2406,4382){\makebox(0,0)[r]{\strut{}transversal}}%
    }%
    \gplgaddtomacro\gplbacktext{%
      \csname LTb\endcsname%
      \put(3354,220){\makebox(0,0){\strut{}L}}%
      \put(4349,220){\makebox(0,0){\strut{}$\Gamma$}}%
      \put(5567,220){\makebox(0,0){\strut{}K}}%
      \put(5699,440){\makebox(0,0)[l]{\strut{} 0}}%
      \put(5699,1018){\makebox(0,0)[l]{\strut{} 2}}%
      \put(5699,1596){\makebox(0,0)[l]{\strut{} 4}}%
      \put(5699,2174){\makebox(0,0)[l]{\strut{} 6}}%
      \put(5699,2752){\makebox(0,0)[l]{\strut{} 8}}%
      \put(5699,3330){\makebox(0,0)[l]{\strut{} 10}}%
      \put(5699,3908){\makebox(0,0)[l]{\strut{} 12}}%
      \put(5699,4486){\makebox(0,0)[l]{\strut{} 14}}%
      \put(6336,2607){\rotatebox{-270}{\makebox(0,0){\strut{}}}}%
      \put(4460,-66){\makebox(0,0){\strut{}}}%
      \put(4460,4665){\makebox(0,0){\strut{}}}%
      \put(4460,4664){\makebox(0,0){\strut{}}}%
      \put(3354,110){\makebox(0,0)[l]{\strut{}}}%
    }%
    \gplgaddtomacro\gplfronttext{%
      \csname LTb\endcsname%
      \put(4957,4602){\makebox(0,0)[r]{\strut{}longitudinal}}%
      \csname LTb\endcsname%
      \put(4957,4382){\makebox(0,0)[r]{\strut{}transversal}}%
    }%
    \gplbacktext
    \put(0,0){\includegraphics[scale=0.8]{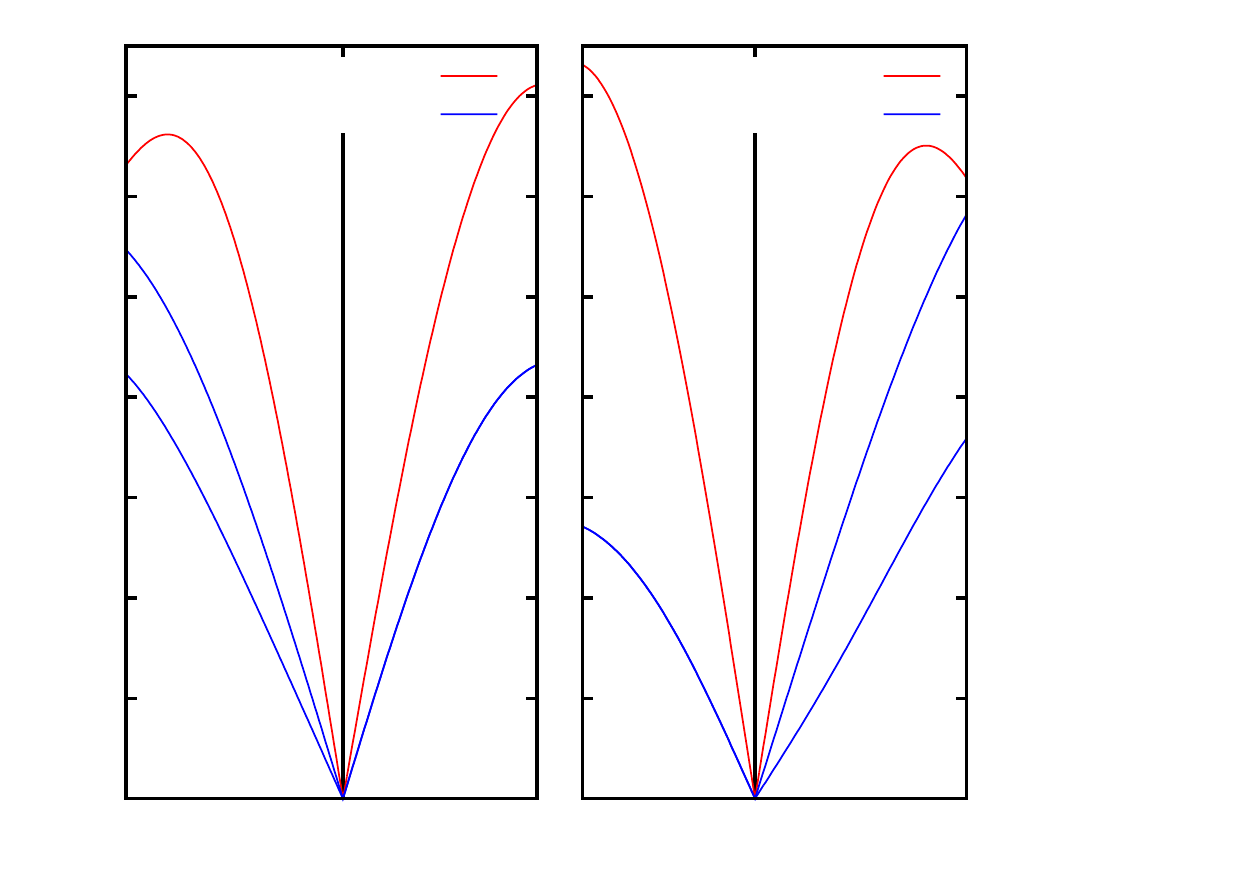}}%
    \gplfronttext
  \end{picture}%
\endgroup

\begingroup
  \makeatletter
  \providecommand\color[2][]{%
    \GenericError{(gnuplot) \space\space\space\@spaces}{%
      Package color not loaded in conjunction with
      terminal option `colourtext'%
    }{See the gnuplot documentation for explanation.%
    }{Either use 'blacktext' in gnuplot or load the package
      color.sty in LaTeX.}%
    \renewcommand\color[2][]{}%
  }%
  \providecommand\includegraphics[2][]{%
    \GenericError{(gnuplot) \space\space\space\@spaces}{%
      Package graphicx or graphics not loaded%
    }{See the gnuplot documentation for explanation.%
    }{The gnuplot epslatex terminal needs graphicx.sty or graphics.sty.}%
    \renewcommand\includegraphics[2][]{}%
  }%
  \providecommand\rotatebox[2]{#2}%
  \@ifundefined{ifGPcolor}{%
    \newif\ifGPcolor
    \GPcolortrue
  }{}%
  \@ifundefined{ifGPblacktext}{%
    \newif\ifGPblacktext
    \GPblacktexttrue
  }{}%
  \let\gplgaddtomacro\g@addto@macro
  \gdef\gplbacktext{}%
  \gdef\gplfronttext{}%
  \makeatother
  \ifGPblacktext
    \def\colorrgb#1{}%
    \def\colorgray#1{}%
  \else
    \ifGPcolor
      \def\colorrgb#1{\color[rgb]{#1}}%
      \def\colorgray#1{\color[gray]{#1}}%
      \expandafter\def\csname LTw\endcsname{\color{white}}%
      \expandafter\def\csname LTb\endcsname{\color{black}}%
      \expandafter\def\csname LTa\endcsname{\color{black}}%
      \expandafter\def\csname LT0\endcsname{\color[rgb]{1,0,0}}%
      \expandafter\def\csname LT1\endcsname{\color[rgb]{0,1,0}}%
      \expandafter\def\csname LT2\endcsname{\color[rgb]{0,0,1}}%
      \expandafter\def\csname LT3\endcsname{\color[rgb]{1,0,1}}%
      \expandafter\def\csname LT4\endcsname{\color[rgb]{0,1,1}}%
      \expandafter\def\csname LT5\endcsname{\color[rgb]{1,1,0}}%
      \expandafter\def\csname LT6\endcsname{\color[rgb]{0,0,0}}%
      \expandafter\def\csname LT7\endcsname{\color[rgb]{1,0.3,0}}%
      \expandafter\def\csname LT8\endcsname{\color[rgb]{0.5,0.5,0.5}}%
    \else
      \def\colorrgb#1{\color{black}}%
      \def\colorgray#1{\color[gray]{#1}}%
      \expandafter\def\csname LTw\endcsname{\color{white}}%
      \expandafter\def\csname LTb\endcsname{\color{black}}%
      \expandafter\def\csname LTa\endcsname{\color{black}}%
      \expandafter\def\csname LT0\endcsname{\color{black}}%
      \expandafter\def\csname LT1\endcsname{\color{black}}%
      \expandafter\def\csname LT2\endcsname{\color{black}}%
      \expandafter\def\csname LT3\endcsname{\color{black}}%
      \expandafter\def\csname LT4\endcsname{\color{black}}%
      \expandafter\def\csname LT5\endcsname{\color{black}}%
      \expandafter\def\csname LT6\endcsname{\color{black}}%
      \expandafter\def\csname LT7\endcsname{\color{black}}%
      \expandafter\def\csname LT8\endcsname{\color{black}}%
    \fi
  \fi
  \setlength{\unitlength}{0.0400bp}%
  \begin{picture}(7200.00,5040.00)%
    \gplgaddtomacro\gplbacktext{%
      \csname LTb\endcsname%
      \put(726,440){\makebox(0,0)[r]{\strut{} 0.3}}%
      \put(726,922){\makebox(0,0)[r]{\strut{} 0.4}}%
      \put(726,1403){\makebox(0,0)[r]{\strut{} 0.5}}%
      \put(726,1885){\makebox(0,0)[r]{\strut{} 0.6}}%
      \put(726,2367){\makebox(0,0)[r]{\strut{} 0.7}}%
      \put(726,2848){\makebox(0,0)[r]{\strut{} 0.8}}%
      \put(726,3330){\makebox(0,0)[r]{\strut{} 0.9}}%
      \put(726,3812){\makebox(0,0)[r]{\strut{} 1}}%
      \put(726,4293){\makebox(0,0)[r]{\strut{} 1.1}}%
      \put(726,4775){\makebox(0,0)[r]{\strut{} 1.2}}%
      \put(858,220){\makebox(0,0){\strut{}W}}%
      \put(2038,220){\makebox(0,0){\strut{}$\Gamma$}}%
      \put(3093,220){\makebox(0,0){\strut{}X}}%
      \put(220,2607){\rotatebox{-270}{\makebox(0,0){\strut{}$\langle\delta n^2\rangle\sigma^3/V$}}}%
      \put(3312,2607){\rotatebox{-270}{\makebox(0,0){\strut{}}}}%
      \put(1975,-66){\makebox(0,0){\strut{}}}%
      \put(1975,4665){\makebox(0,0){\strut{}}}%
      \put(1975,4664){\makebox(0,0){\strut{}}}%
      \put(66,110){\makebox(0,0)[l]{\strut{}}}%
    }%
    \gplgaddtomacro\gplfronttext{%
      \csname LTb\endcsname%
      \put(2011,2367){\makebox(0,0){\strut{}$q\sigma$}}%
    }%
    \gplgaddtomacro\gplbacktext{%
      \csname LTb\endcsname%
      \put(1301,2708){\makebox(0,0)[r]{\strut{}6}}%
      \put(1301,3061){\makebox(0,0)[r]{\strut{}9}}%
      \put(1301,3414){\makebox(0,0)[r]{\strut{}12}}%
      \put(1301,3767){\makebox(0,0)[r]{\strut{}15}}%
      \put(1328,2554){\makebox(0,0){\strut{}0.01}}%
      \put(2038,2554){\makebox(0,0){\strut{}\large$\longleftrightarrow$}}%
      \put(2747,2554){\makebox(0,0){\strut{}0.01}}%
      \put(2966,3237){\rotatebox{-270}{\makebox(0,0){\strut{}}}}%
      \put(2037,2202){\makebox(0,0){\strut{}}}%
      \put(2037,3657){\makebox(0,0){\strut{}}}%
      \put(2037,3656){\makebox(0,0){\strut{}}}%
      \put(800,2378){\makebox(0,0)[l]{\strut{}}}%
    }%
    \gplgaddtomacro\gplfronttext{%
    }%
    \gplgaddtomacro\gplbacktext{%
      \csname LTb\endcsname%
      \put(3656,2554){\makebox(0,0){\strut{}0.01}}%
      \put(4289,2554){\makebox(0,0){\strut{}\large$\longleftrightarrow$}}%
      \put(4921,2554){\makebox(0,0){\strut{}0.01}}%
      \put(4948,2708){\makebox(0,0)[l]{\strut{}6}}%
      \put(4948,3061){\makebox(0,0)[l]{\strut{}9}}%
      \put(4948,3414){\makebox(0,0)[l]{\strut{}12}}%
      \put(4948,3767){\makebox(0,0)[l]{\strut{}15}}%
      \put(5558,3237){\rotatebox{-270}{\makebox(0,0){\strut{}}}}%
      \put(4288,2202){\makebox(0,0){\strut{}}}%
      \put(4288,3657){\makebox(0,0){\strut{}}}%
      \put(4288,3656){\makebox(0,0){\strut{}}}%
      \put(3656,2378){\makebox(0,0)[l]{\strut{}}}%
    }%
    \gplgaddtomacro\gplfronttext{%
    }%
    \gplgaddtomacro\gplbacktext{%
      \csname LTb\endcsname%
      \put(3354,220){\makebox(0,0){\strut{}L}}%
      \put(4289,220){\makebox(0,0){\strut{}$\Gamma$}}%
      \put(5435,220){\makebox(0,0){\strut{}K}}%
      \put(5462,440){\makebox(0,0)[l]{\strut{} 0.3}}%
      \put(5462,922){\makebox(0,0)[l]{\strut{} 0.4}}%
      \put(5462,1403){\makebox(0,0)[l]{\strut{} 0.5}}%
      \put(5462,1885){\makebox(0,0)[l]{\strut{} 0.6}}%
      \put(5462,2367){\makebox(0,0)[l]{\strut{} 0.7}}%
      \put(5462,2848){\makebox(0,0)[l]{\strut{} 0.8}}%
      \put(5462,3330){\makebox(0,0)[l]{\strut{} 0.9}}%
      \put(5462,3812){\makebox(0,0)[l]{\strut{} 1}}%
      \put(5462,4293){\makebox(0,0)[l]{\strut{} 1.1}}%
      \put(5462,4775){\makebox(0,0)[l]{\strut{} 1.2}}%
      \put(6336,2607){\rotatebox{-270}{\makebox(0,0){\strut{}}}}%
      \put(4394,-66){\makebox(0,0){\strut{}}}%
      \put(4394,4665){\makebox(0,0){\strut{}}}%
      \put(4394,4664){\makebox(0,0){\strut{}}}%
      \put(3354,110){\makebox(0,0)[l]{\strut{}}}%
    }%
    \gplgaddtomacro\gplfronttext{%
      \csname LTb\endcsname%
      \put(2435,4555){\makebox(0,0)[r]{\strut{}$\nu^{-1}k_BTn^2_0\sigma^3$}}%
      \csname LTb\endcsname%
      \put(2435,4115){\makebox(0,0)[r]{\strut{}$\langle\delta n^2\rangle\sigma^3/V$}}%
      \csname LTb\endcsname%
      \put(4610,4555){\makebox(0,0)[r]{\strut{}$10^{10}\Delta$}}%
      \csname LTb\endcsname%
      \put(4256,2367){\makebox(0,0){\strut{}$q\sigma$}}%
    }%
    \gplbacktext
    \put(0,0){\includegraphics[scale=0.8]{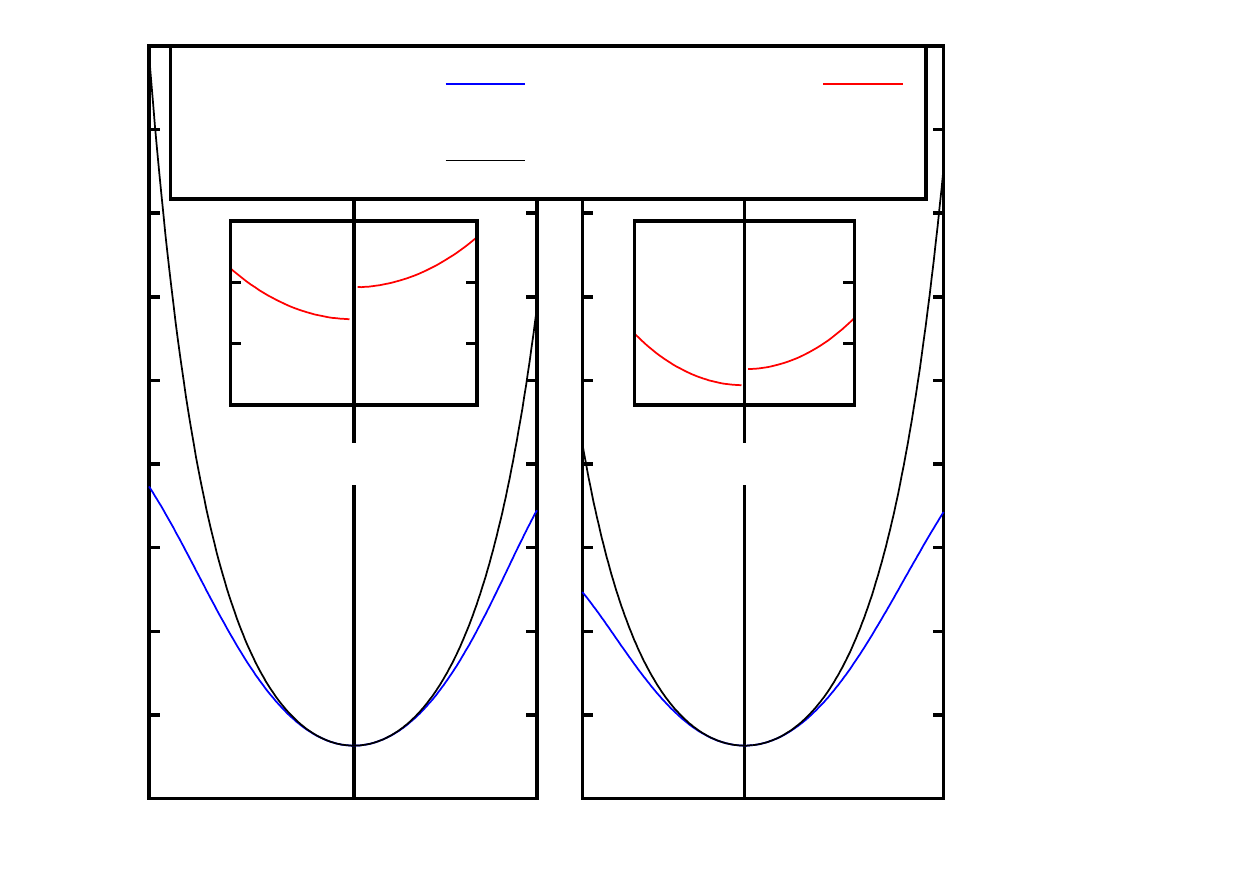}}%
    \gplfronttext
  \end{picture}%
\endgroup

\caption{Top panel: Phonon dispersion relation for a cluster crystal with fcc structure along four symmetry lines in the first Brillouin zone; the state at $k_BT/\epsilon=1.1$ and $n_0\sigma^3=8.5$ is also included in Figs.~\ref{fig2} and \ref{fig3} and Table I.\\ Bottom panel: The $\mathbf{q}$-dependent density correlation function from Eq.~\eqref{eq57} and its dominating part for small $q$ given by $\nu^{-1}(\mathbf{q})$ 
 (blue). Insets: The difference of both quantities $\Delta(\mathbf{q})=\langle\delta n^2(\mathbf{q})\rangle\sigma^3/V-\nu^{-1}(\mathbf{q})k_BTn^2_0$ in a small range around $\mathbf{q}=0$ for the same symmetry lines. The different limits  $\Delta({\bf q}\to 0)$ depending on direction are apparent.}\label{fig4n}
\end{figure}

The correlation functions for the coarse-grained fields can be obtained from the $\bf q$-dependent elastic coefficients according to Eq.~(\ref{InverseAll}). They follow from the density profile obtained by minimizing the DFT free energy functional. The top panel in Fig.~\ref{fig4n} shows the dispersion relations obtained from diagonalizing the dynamic matrix appearing in the wave-equation of the momentum density \cite{Walz10}: $D_{\alpha\beta}(\mathbf{q}) =\Lambda_{\alpha\beta}({\bf q})/(mn_0)$ with particle mass $m$, and $\mathbf \Lambda$ given in Eq.~\eqref{eq19}. A typical state with fcc lattice is considered. The eigenfrequencies $\omega$ of $D_{\alpha\beta}$ exhibit the familiar longitudinal and (up to two) transversal acoustic branches depending on the $\bf q$-directions followed in the first Brillouin zone. Remarkably, the high degree of disorder contained in the broad distributions of occupation numbers does not weaken the solid overly; the dispersion relations exhibit the shapes familiar from ideal solids and assume magnitudes comparable to the values obtained from potential expansions at $T=0$ assuming ideality \cite{Likos07}.  

While the direction-dependence of the dispersion relations is familiar, the  concomitant direction dependence of the density correlation functions had not been established. The lower panel in Fig.~\ref{fig4n} shows the density correlation function $\langle \delta n^\ast({\bf q}) \delta n({\bf q}) \rangle$ from Eq.~\eqref{PerKorrN} and $\nu({\bf q})$ from Eq. (\ref{numulambda}a). The latter is the $q$-dependent generalization of the thermodynamic derivative $\nu=n_0^2 (\partial \mu/\partial n)_{u_{\alpha\beta}}$ from Eq.~\eqref{thermoderiv}. Both functions almost completely agree for small wavevectors because of the extremely weak coupling between density and strain in cluster crystals; the coefficient $\partial^2 f/\partial n \partial u_{\alpha\beta}=\mu_{\alpha\beta}=\mu_0\delta_{\alpha\beta}$ from Eq.~(\ref{thermoderiv}b), which is diagonal in fcc lattices, is very small: $\mu_0/\nu\approx 2\cdot 10^{-4}$. Both functions start deviating for wavevectors approaching the Brillouin zone boundary.  Because $\nu({\bf q})$ possesses a regular small $\bf q$ expansion given in  Eq. (\ref{numulambda}b), the non-analyticity of the density correlation function can be brought out by considering the difference $\Delta(\mathbf{q})=\langle\delta n^2(\mathbf{q})\rangle\sigma^3/V-\nu^{-1}(\mathbf{q})k_BTn^2_0$. This $\Delta$ is small for small wavevectors because the small factor $\mu_0$ enters quadratically. Yet, it clearly shows different limits for ${\bf q}\to0$ resulting from the direction dependence discussed in context with Eq.~\eqref{dyncomp}. The insets in  Fig.~\ref{fig4n} show the curves obtained from taking the limit ${\bf q}\to0$ along high-symmetry directions in the first Brillouin zone of an fcc cluster crystal. The directions go from the center $\Gamma$ of the Brillouin zone
 along direction $[120]$ (given by Miller indices \cite{Ashcroft76}) to the point W, along $[010]$ to X, along $[111]$ to L, and along $[110]$ to K. Along each of these directions, the density correlation function  $\langle \delta n^\ast({\bf q}) \delta n({\bf q}) \rangle$ takes a different limit for ${\bf q}\to0$. The very small magnitude of the differences results from the small value of $\mu_0/\nu$ specific to cluster crystals; the differences are numerically reliable. 
 
\subsection{Discussion of low temperature phase transitions}

\begin{figure}[tb]
 \centering
\begingroup
  \makeatletter
  \providecommand\color[2][]{%
    \GenericError{(gnuplot) \space\space\space\@spaces}{%
      Package color not loaded in conjunction with
      terminal option `colourtext'%
    }{See the gnuplot documentation for explanation.%
    }{Either use 'blacktext' in gnuplot or load the package
      color.sty in LaTeX.}%
    \renewcommand\color[2][]{}%
  }%
  \providecommand\includegraphics[2][]{%
    \GenericError{(gnuplot) \space\space\space\@spaces}{%
      Package graphicx or graphics not loaded%
    }{See the gnuplot documentation for explanation.%
    }{The gnuplot epslatex terminal needs graphicx.sty or graphics.sty.}%
    \renewcommand\includegraphics[2][]{}%
  }%
  \providecommand\rotatebox[2]{#2}%
  \@ifundefined{ifGPcolor}{%
    \newif\ifGPcolor
    \GPcolortrue
  }{}%
  \@ifundefined{ifGPblacktext}{%
    \newif\ifGPblacktext
    \GPblacktexttrue
  }{}%
  \let\gplgaddtomacro\g@addto@macro
  \gdef\gplbacktext{}%
  \gdef\gplfronttext{}%
  \makeatother
  \ifGPblacktext
    \def\colorrgb#1{}%
    \def\colorgray#1{}%
  \else
    \ifGPcolor
      \def\colorrgb#1{\color[rgb]{#1}}%
      \def\colorgray#1{\color[gray]{#1}}%
      \expandafter\def\csname LTw\endcsname{\color{white}}%
      \expandafter\def\csname LTb\endcsname{\color{black}}%
      \expandafter\def\csname LTa\endcsname{\color{black}}%
      \expandafter\def\csname LT0\endcsname{\color[rgb]{1,0,0}}%
      \expandafter\def\csname LT1\endcsname{\color[rgb]{0,1,0}}%
      \expandafter\def\csname LT2\endcsname{\color[rgb]{0,0,1}}%
      \expandafter\def\csname LT3\endcsname{\color[rgb]{1,0,1}}%
      \expandafter\def\csname LT4\endcsname{\color[rgb]{0,1,1}}%
      \expandafter\def\csname LT5\endcsname{\color[rgb]{1,1,0}}%
      \expandafter\def\csname LT6\endcsname{\color[rgb]{0,0,0}}%
      \expandafter\def\csname LT7\endcsname{\color[rgb]{1,0.3,0}}%
      \expandafter\def\csname LT8\endcsname{\color[rgb]{0.5,0.5,0.5}}%
    \else
      \def\colorrgb#1{\color{black}}%
      \def\colorgray#1{\color[gray]{#1}}%
      \expandafter\def\csname LTw\endcsname{\color{white}}%
      \expandafter\def\csname LTb\endcsname{\color{black}}%
      \expandafter\def\csname LTa\endcsname{\color{black}}%
      \expandafter\def\csname LT0\endcsname{\color{black}}%
      \expandafter\def\csname LT1\endcsname{\color{black}}%
      \expandafter\def\csname LT2\endcsname{\color{black}}%
      \expandafter\def\csname LT3\endcsname{\color{black}}%
      \expandafter\def\csname LT4\endcsname{\color{black}}%
      \expandafter\def\csname LT5\endcsname{\color{black}}%
      \expandafter\def\csname LT6\endcsname{\color{black}}%
      \expandafter\def\csname LT7\endcsname{\color{black}}%
      \expandafter\def\csname LT8\endcsname{\color{black}}%
    \fi
  \fi
  \setlength{\unitlength}{0.03500bp}%
  \begin{picture}(7200.00,5040.00)%
    \gplgaddtomacro\gplbacktext{%
      \csname LTb\endcsname%
      \put(946,704){\makebox(0,0)[r]{\strut{}0}}%
      \put(946,1286){\makebox(0,0)[r]{\strut{}0.01}}%
      \put(946,1867){\makebox(0,0)[r]{\strut{}0.02}}%
      \put(946,2449){\makebox(0,0)[r]{\strut{}0.03}}%
      \put(946,3030){\makebox(0,0)[r]{\strut{}0.04}}%
      \put(946,3612){\makebox(0,0)[r]{\strut{}0.05}}%
      \put(946,4193){\makebox(0,0)[r]{\strut{}0.06}}%
      \put(946,4775){\makebox(0,0)[r]{\strut{}0.07}}%
      \put(1078,484){\makebox(0,0){\strut{} 1}}%
      \put(2482,484){\makebox(0,0){\strut{} 1.5}}%
      \put(3886,484){\makebox(0,0){\strut{} 2}}%
      \put(5289,484){\makebox(0,0){\strut{} 2.5}}%
      \put(6693,484){\makebox(0,0){\strut{} 3}}%
      \put(176,2739){\rotatebox{-270}{\makebox(0,0){\strut{}$k_BT/\epsilon$}}}%
      \put(6912,2739){\rotatebox{-270}{\makebox(0,0){\strut{}}}}%
      \put(3885,154){\makebox(0,0){\strut{}$n_0\sigma^3$}}%
      \put(3885,4665){\makebox(0,0){\strut{}}}%
      \put(3885,4664){\makebox(0,0){\strut{}}}%
      \put(286,110){\makebox(0,0)[l]{\strut{}}}%
      \put(3105,4143){\makebox(0,0)[l]{\strut{}$\sqrt{\langle\Delta n_c^2\rangle}=0.3$}}%
	  \put(2482,1639){\rotatebox{-270}{\makebox(0,0){\strut{}fcc3}}}%
	  \put(3886,1639){\rotatebox{-270}{\makebox(0,0){\strut{}fcc4}}}%
	  \put(5289,1639){\rotatebox{-270}{\makebox(0,0){\strut{}fcc5}}}%
    }%
    \gplgaddtomacro\gplfronttext{%
    }%
    \gplbacktext
    \put(0,0){\includegraphics[scale=0.7]{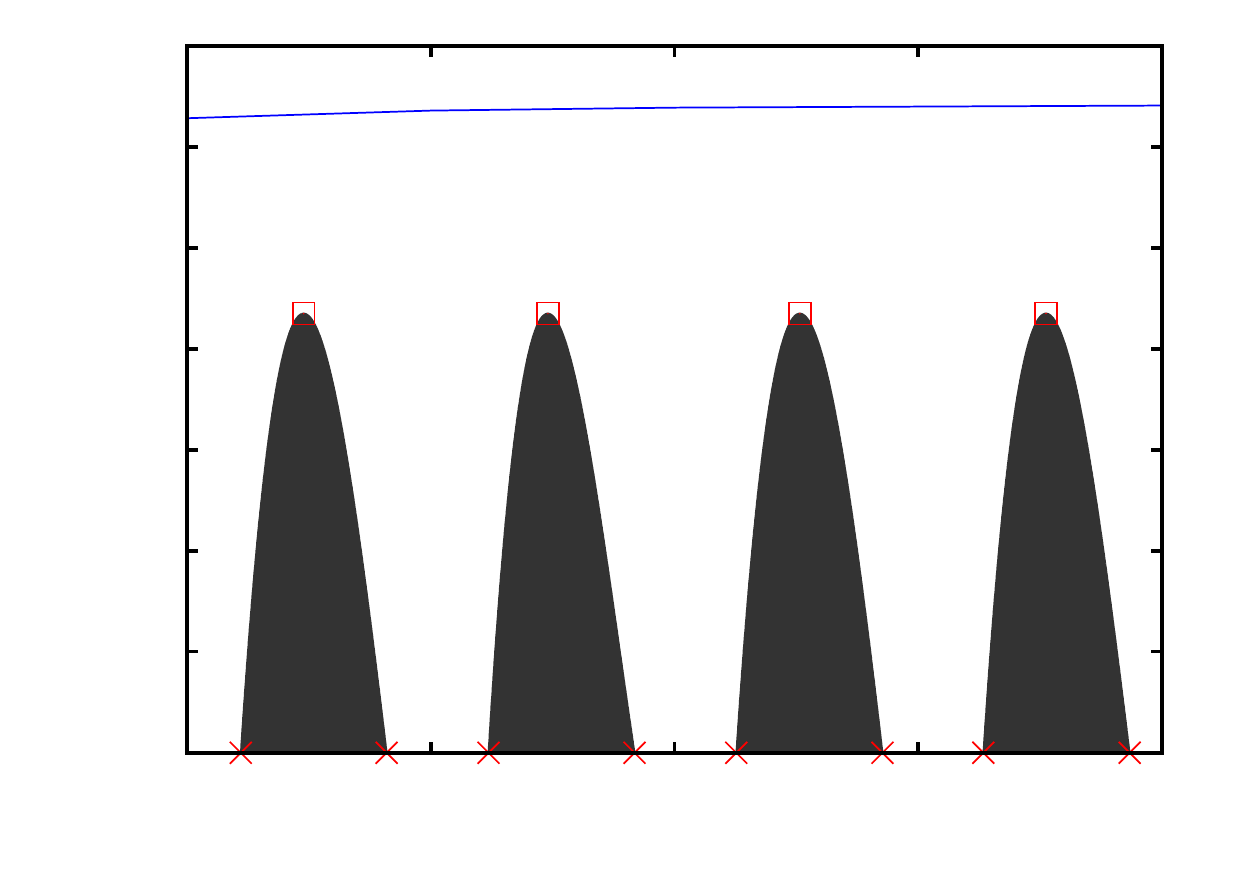}}%
    \gplfronttext
  \end{picture}%
\endgroup

 \caption{Low-temperature phase diagram of the GEM-4 system as determined in MC simulations\cite{Wilding13}; red data points connected by lines as guides to the eye indicate the coexistence regions. Pure fcc phases with integer site-occupations (denoted fccn with $n=2,3,\ldots$) survive only at extremely low temperatures. Mean-field DFT  provides a good estimates of the critical temperatures for a reasonable numerical value of the occupation number variance,  $\sqrt{\langle\Delta n_c^2\rangle}=0.3$ (blue line). }
 \label{fig4}
\end{figure}

Figure~\ref{fig1} shows only little variation of $\kappa^c n^2_0$, especially in the low temperature/high density range. Because $\frac{n_c}{n_0}$
also varies little, the variance of the occupation number fluctuations, $\langle\Delta n_c^2\rangle$, is nearly independent of the density and scales mainly with the temperature. This points to an internal inconsistency of the mean-field description at low temperatures. The width of the occupation number distributions vanishes for $T\to0$, yet, non-integer average occupation numbers can occur. The failure to find integer occupations clearly indicates the break-down of mean-field theory for low temperatures. Simulations  show that
the phase diagram of the GEM-4-system exhibits fcc phases where the occupation numbers take integer values at low temperatures \cite{Zhang12,Wilding13}.  Phase coexistence regions lie between them; see Fig.~\ref{fig4} showing simulations from Ref.~[\onlinecite{Wilding13}]. At  critical temperatures each coexistence region vanishes, and the homogeneous fcc phase with a distribution of occupations becomes stable. The MC simulations\cite{Wilding13} suggest that these critical temperatures are nearly identical for each phase coexistence, i.e they are nearly
independent of the density. The mean field density functional approach only describes the homogeneous distributed phase and misses the coexistence regions at low temperatures. Potential energy minimization at zero temperature gives homogeneous integer occupations and rationalizes their coexistences \cite{Likos07}.   

Still, the knowledge of the occupation number fluctuations in the homogeneous phase allows to establish a criterion when the homogeneous phase is not consistent.  We suggest that there is a threshold 
of the occupation number variance $\langle\Delta n_c^2\rangle$ where the hopping between the lattice sites becomes strong enough to lift the (zero temperature) restriction of integer occupation numbers.  
Consequently, for temperatures below this value, we expect the mean field density functional \eqref{df} to break down and integer occupation phases to become stable, as holds at zero temperature. Figure \ref{fig4} shows that the estimate of the occupation number deviation $\sqrt{\langle\Delta n_c^2\rangle}=0.3$ gives an order of magnitude estimate of the critical temperatures. 

\begin{figure}[htb]
\begingroup
  \makeatletter
  \providecommand\color[2][]{%
    \GenericError{(gnuplot) \space\space\space\@spaces}{%
      Package color not loaded in conjunction with
      terminal option `colourtext'%
    }{See the gnuplot documentation for explanation.%
    }{Either use 'blacktext' in gnuplot or load the package
      color.sty in LaTeX.}%
    \renewcommand\color[2][]{}%
  }%
  \providecommand\includegraphics[2][]{%
    \GenericError{(gnuplot) \space\space\space\@spaces}{%
      Package graphicx or graphics not loaded%
    }{See the gnuplot documentation for explanation.%
    }{The gnuplot epslatex terminal needs graphicx.sty or graphics.sty.}%
    \renewcommand\includegraphics[2][]{}%
  }%
  \providecommand\rotatebox[2]{#2}%
  \@ifundefined{ifGPcolor}{%
    \newif\ifGPcolor
    \GPcolortrue
  }{}%
  \@ifundefined{ifGPblacktext}{%
    \newif\ifGPblacktext
    \GPblacktexttrue
  }{}%
  \let\gplgaddtomacro\g@addto@macro
  \gdef\gplbacktext{}%
  \gdef\gplfronttext{}%
  \makeatother
  \ifGPblacktext
    \def\colorrgb#1{}%
    \def\colorgray#1{}%
  \else
    \ifGPcolor
      \def\colorrgb#1{\color[rgb]{#1}}%
      \def\colorgray#1{\color[gray]{#1}}%
      \expandafter\def\csname LTw\endcsname{\color{white}}%
      \expandafter\def\csname LTb\endcsname{\color{black}}%
      \expandafter\def\csname LTa\endcsname{\color{black}}%
      \expandafter\def\csname LT0\endcsname{\color[rgb]{1,0,0}}%
      \expandafter\def\csname LT1\endcsname{\color[rgb]{0,1,0}}%
      \expandafter\def\csname LT2\endcsname{\color[rgb]{0,0,1}}%
      \expandafter\def\csname LT3\endcsname{\color[rgb]{1,0,1}}%
      \expandafter\def\csname LT4\endcsname{\color[rgb]{0,1,1}}%
      \expandafter\def\csname LT5\endcsname{\color[rgb]{1,1,0}}%
      \expandafter\def\csname LT6\endcsname{\color[rgb]{0,0,0}}%
      \expandafter\def\csname LT7\endcsname{\color[rgb]{1,0.3,0}}%
      \expandafter\def\csname LT8\endcsname{\color[rgb]{0.5,0.5,0.5}}%
    \else
      \def\colorrgb#1{\color{black}}%
      \def\colorgray#1{\color[gray]{#1}}%
      \expandafter\def\csname LTw\endcsname{\color{white}}%
      \expandafter\def\csname LTb\endcsname{\color{black}}%
      \expandafter\def\csname LTa\endcsname{\color{black}}%
      \expandafter\def\csname LT0\endcsname{\color{black}}%
      \expandafter\def\csname LT1\endcsname{\color{black}}%
      \expandafter\def\csname LT2\endcsname{\color{black}}%
      \expandafter\def\csname LT3\endcsname{\color{black}}%
      \expandafter\def\csname LT4\endcsname{\color{black}}%
      \expandafter\def\csname LT5\endcsname{\color{black}}%
      \expandafter\def\csname LT6\endcsname{\color{black}}%
      \expandafter\def\csname LT7\endcsname{\color{black}}%
      \expandafter\def\csname LT8\endcsname{\color{black}}%
    \fi
  \fi
  \setlength{\unitlength}{0.03500bp}%
  \begin{picture}(7200.00,5040.00)%
    \gplgaddtomacro\gplbacktext{%
      \csname LTb\endcsname%
      \put(946,704){\makebox(0,0)[r]{\strut{} 0}}%
      \put(946,1518){\makebox(0,0)[r]{\strut{} 0.1}}%
      \put(946,2332){\makebox(0,0)[r]{\strut{} 0.2}}%
      \put(946,3147){\makebox(0,0)[r]{\strut{} 0.3}}%
      \put(946,3961){\makebox(0,0)[r]{\strut{} 0.4}}%
      \put(946,4775){\makebox(0,0)[r]{\strut{} 0.5}}%
      \put(1078,484){\makebox(0,0){\strut{} 0}}%
      \put(1702,484){\makebox(0,0){\strut{} 0.02}}%
      \put(2326,484){\makebox(0,0){\strut{} 0.04}}%
      \put(2950,484){\makebox(0,0){\strut{} 0.06}}%
      \put(3574,484){\makebox(0,0){\strut{} 0.08}}%
      \put(4197,484){\makebox(0,0){\strut{} 0.1}}%
      \put(4821,484){\makebox(0,0){\strut{} 0.12}}%
      \put(5445,484){\makebox(0,0){\strut{} 0.14}}%
      \put(6069,484){\makebox(0,0){\strut{} 0.16}}%
      \put(6693,484){\makebox(0,0){\strut{} 0.18}}%
      \put(176,2739){\rotatebox{-270}{\makebox(0,0){\strut{}$\sqrt{\Delta n_c^2}$}}}%
      \put(6912,2739){\rotatebox{-270}{\makebox(0,0){\strut{}}}}%
      \put(4017,154){\makebox(0,0){\strut{}$k_B T/\epsilon$}}%
      \put(3885,4665){\makebox(0,0){\strut{}}}%
      \put(3885,4664){\makebox(0,0){\strut{}}}%
      \put(286,110){\makebox(0,0)[l]{\strut{}}}%
    }%
    \gplgaddtomacro\gplfronttext{%
      \csname LTb\endcsname%
      \put(3921,3039){\makebox(0,0)[l]{\strut{}$n_c=3, n_0\sigma^3=1.5$}}%
      \csname LTb\endcsname%
      \put(3921,2646){\makebox(0,0)[l]{\strut{}$n_c=4, n_0\sigma^3=2$}}%
      \csname LTb\endcsname%
      \put(3921,2253){\makebox(0,0)[l]{\strut{}$n_c=5, n_0\sigma^3=2.5$}}%
      \csname LTb\endcsname%
      \put(3921,1860){\makebox(0,0)[l]{\strut{}$n_c=6, n_0\sigma^3=3$}}%
      \csname LTb\endcsname%
      \put(3921,1467){\makebox(0,0)[l]{\strut{}$k_BT/\epsilon=0.0471$}}%
      \csname LTb\endcsname%
      \put(3921,1074){\makebox(0,0)[l]{\strut{}$\sqrt{\langle\Delta n_c^2\rangle}=0.3$}}%
    }%
    \gplbacktext
    \put(0,0){\includegraphics[scale=0.7]{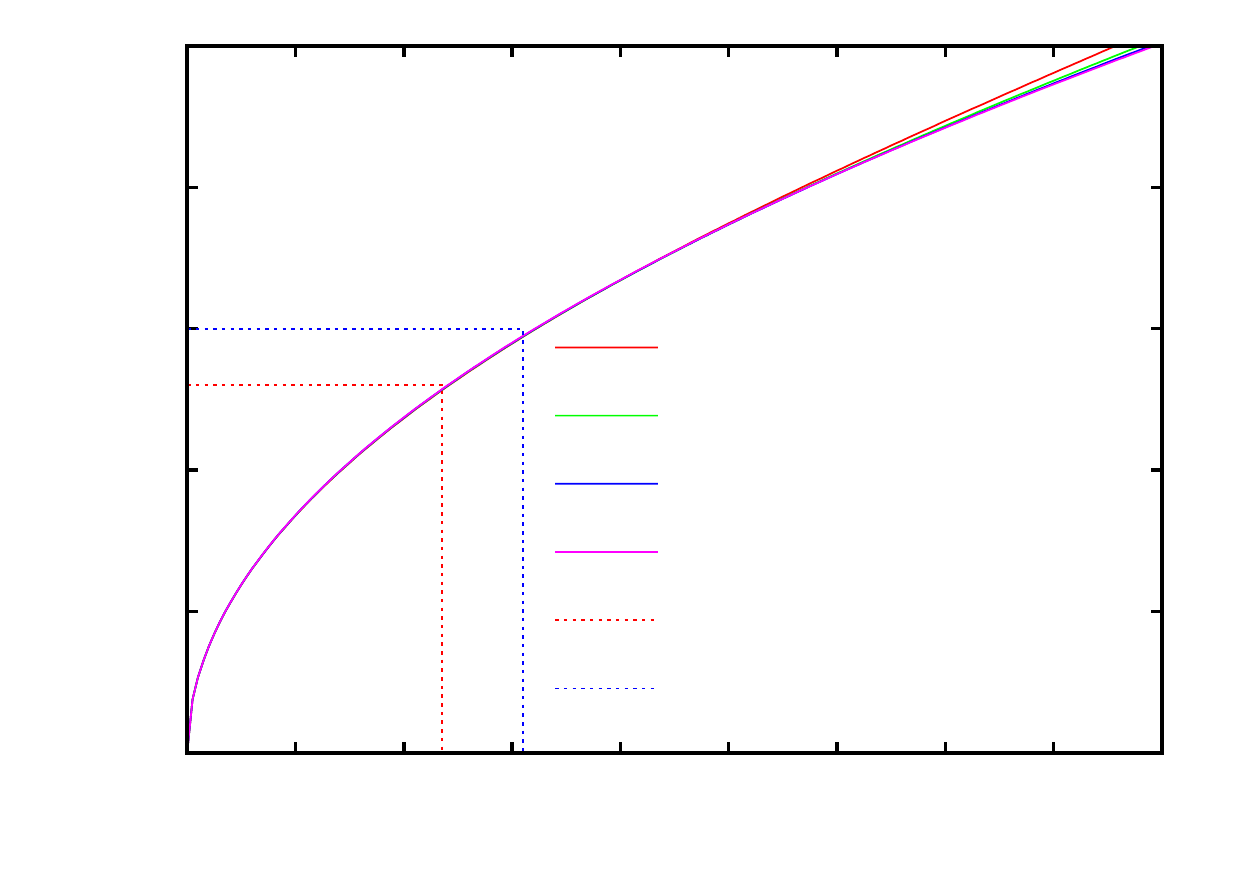}}%
    \gplfronttext
  \end{picture}%
\endgroup

\caption{The occupation number fluctuation $\sqrt{<\Delta n_{c}^{2}>}$ (standard deviation) versus the temperature in units of $\epsilon/k_{B}$. The standard deviation\rem{$\sqrt{<\Delta n_{c}^{2}>}$}
is a function of temperature and density. It is plotted for several integer occupied states. The densities are chosen with the approximation $n_c/n_0\approx 2$, which differs by about two per cent from the optimal DFT-value. The red dotted line denotes the point of the curve at the critical temperature $k_{B}T_{c}/\epsilon=0.471$ which is obtained from [N]pT simulations\cite{Zhang12}, the blue line the estimate of Fig.~\ref{fig4}.} \label{fig5}
\end{figure}

The adequacy of the suggested criterion and the stability of the estimate can be studied in a little more detail.
Figure~\ref{fig5} shows the occupation number fluctuation for several phases with integer occupations as function of temperature. Here integer occupation numbers were enforced by hand before minimizing the functional in Eq.~\eqref{df} by varying $\alpha$ only. The critical temperatures observed in simulations are well compatible with an occupation number variation of  $\sqrt{\langle\Delta n_c^2\rangle}\approx 0.25$, which appears a rather reasonable value  for enabling hopping to smear out the occupation numbers on different lattice sites. Moreover, varying the value of this criterion moves the estimates of the critical temperatures only little. For different integer occupations, they differ only slightly.  

\section{Conclusions and outlook}

We derived the thermodynamic expression for the isothermal compressibility $\kappa$ in a general crystal, and discussed its connection to the small wavevector limit of the density correlation function. The correlation functions of coarse grained fields of macroscopic elasticity theory were calculated within the framework of density functional theory, allowing for a finite  density of defects. Explicit expressions for the coefficients in the phenomenological free energy in terms of the direct correlation function of density functional theory were obtained. The correlation function of the coarse-grained density field does not, in general, determine the compressibility. For the case of an ideal isotropic solid, we could identify the origin of the discrepancy from a calculation in macroscopic elasticity theory. It arises from the long-ranged strain fluctuations which decay like $1/r^3$ and thereby cause boundary effects to enter the elastic energy. While in systems with spontaneously broken symmetry, anomalous longitudinal correlations exist in general\cite{Zwerger04} (besides the familiar symmetry restoring fluctuations\cite{Forster75}), the present observation appears more related to long-ranged dipolar correlations in polar fluids\cite{Deutch73}. There, the dielectric tensor in response to the vacuum electric field depends on the shape of the material and on the boundary conditions. It can be connected to a well-defined isotropic dielectric constant only via shape/ boundary-effect dependent distribution functions. To work out a corresponding relation for arbitrary symmetries and sample shapes of crystalline solids is left for future work.

We applied the theory to the elasticity of cluster crystals made by soft particles. In these crystals, the fluctuating occupation numbers of lattice sites play the role of local defects and strongly affect the stable phases and their material responses. Therefore, cluster crystals appear an ideal system to test our theory. The obtained compressibilities and occupation number distributions compare well with data obtained in Monte Carlo simulations. Mean-field theory breaks down at low temperatures. Yet, the theory can be used in order to identify the temperature range where mean-field theory breaks down. This provides rather reasonable and stable estimates for the critical temperatures, below which the zero-temperature phases with integer occupation numbers are stable.

\begin{acknowledgments}
We thank Florian Miserez and Tadeus Ras for useful discussions.  This work
was partially funded by the German Excellence Initiative (CW \& MF). The work was started when GS visited Konstanz, which was made possible by Humboldt Foundation and Zukunftskolleg of Universit\"at Konstanz. Partial support by NSF Grants CHE 0909676 and CHE 1213401 is gratefully acknowledged (GS).
\end{acknowledgments}

\appendix
\section{Thermodynamic manipulations} \label{appComp}

As a consequence of \eqref{firstlawF} a Gibbs-Duhem relation can be derived
\begin{equation}\label{gibbsduhem}
-Vdp+Nd\mu+U_{\alpha\beta}dh_{\alpha\beta}=0. 
\end{equation}
It states that the pressure obeys $p=p(\mu,h_{\alpha\beta})$, which can be used to simplify the total differential of the free energy density per volume $f=F/V=\mu n - p + u_{\alpha\beta}h_{\alpha\beta}$. It is a proper density because the free energy $F$ is a homogeneous function of its extensive variables. As a first result, from the Gibbs-Duhem relation \eqref{gibbsduhem}, the total differential of $f$ given in Eq.~\eqref{dfeq} follows. Also Eq.~\eqref{gibbsduhem}
 yields for an isothermal change with $dh_{\alpha\beta}=0$
\begin{equation}
Nd\mu = Vdp,
\end{equation}
which can be used for 
\begin{align}\label{AppendixKappa}
\frac{1}{N}\frac{\partial N}{\partial \mu}\Big|_{V,h_{\alpha\beta}}&=\frac{1}{V}\frac{\partial N}{\partial p}\Big|_{V,h_{\alpha\beta}}=\frac{\partial n}{\partial p}\Big|_{V,h_{\alpha\beta}}\nonumber\\
&=-\frac{n_0}{V}\frac{\partial V}{\partial p}\Big|_{N,h_{\alpha\beta}}=n_0\kappa.
\end{align}
This verifies Eq.~\eqref{compdef} as the compressibility at constant stress tensor $h_{\alpha\beta}$. 
%

\subsection{Alternative formula for $\kappa$ and $\kappa^c$}

The discussion of  the quadratic terms in the free energy can be related to more standard thermodynamic considerations, which provides additional support for our results. To find
 an alternative formula for the compressibility at constant strain $h_{\alpha\beta}$, we start with Eq.~\eqref{AppendixKappa} and assume a relation $\mathbf{u}(\mathbf{h},\mu)$
\begin{align}
\kappa &= \frac{1}{n_0^2}\frac{\partial n}{\partial \mu}\Big|_{h_{\alpha\beta}}=
\frac{1}{n_0^2}\frac{\partial n}{\partial \mu}\Big|_{u_{\alpha\beta}}+\frac{1}{n_0^2}\frac{\partial n}{\partial u_{\alpha\beta}}\Big|_{\mu}\frac{\partial u_{\alpha\beta}}{\partial \mu}\Big|_{h_{\alpha\beta}}
\end{align}
The last derivative is at constant $h_{\alpha\beta}$. With
\begin{equation}
0=dh_{\alpha\beta}=\frac{\partial h_{\gamma\delta}}{\partial u_{\alpha\beta}}\Big|_\mu du_{\alpha\beta}+\frac{\partial h_{\gamma\delta}}{\partial \mu}d\mu
\end{equation} 
it can be written as
\begin{equation}
\frac{\partial u_{\alpha\beta}}{\partial \mu}\Big|_{h_{\alpha\beta}}=- \left(\frac{\partial h_{\gamma\delta}}{\partial u_{\alpha\beta}}\Big|_\mu\right)^{-1}\frac{\partial h_{\gamma\delta}}{\partial \mu}\Big|_{u_{\alpha\beta}}.
\end{equation}
which leads to the alternative formula
\begin{equation}\label{kappa_alt}
\kappa=\frac{1}{n_0^2}\frac{\partial n}{\partial \mu}\Big|_{u_{\alpha\beta}}-\frac{1}{n_0^2}\frac{\partial n}{\partial u_{\alpha\beta}}\Big|_{\mu}\left(\frac{\partial h_{\gamma\delta}}{\partial u_{\alpha\beta}}\Big|_\mu\right)^{-1}\frac{\partial h_{\gamma\delta}}{\partial \mu}\Big|_{u_{\alpha\beta}}.
\end{equation}
The inverse of 
\begin{equation}
\frac{\partial^2 f}{\partial u_{\alpha\beta}u_{\gamma\delta}}
=C_{\alpha\beta\gamma\delta}\nonumber
\end{equation} is defined by\cite{Wallace70}
\begin{align}
C_{\alpha\beta\gamma\delta}C^{-1}_{\gamma\delta\mu\nu} &=\frac{1}{2}( \delta_{\alpha\mu}\delta_{\beta\nu}+\delta_{\alpha\nu}\delta_{\beta\mu}).
\end{align}
The unusual definition for the "unit matrix" is a consequence from the symmetrisation of the strain tensor, $u_{\alpha\beta}=\frac{1}{2}(\nabla_{\alpha}u_{\beta}+\nabla_{\beta}u_{\alpha})$, and holds for all second order derivatives with respect to $u_{\alpha\beta}$.

The thermodynamic derivatives can be expressed through the elastic coefficients  $\nu,\mu_{\alpha\beta},C_{\alpha\beta\gamma\delta}$ as follows:
The first term of the compressibility is basically the only non-vanishing term in the fluid limit
\begin{equation}
\frac{1}{n_0^2}\frac{\partial n}{\partial \mu}\Big|_{u_{\alpha\beta}} = \Big(n_0^2\frac{\partial \mu}{\partial n}\Big|_{u_{\alpha\beta}}\Big)^{-1}  = \nu^{-1}.
\end{equation}
For the second term the chemical potential $\mu$ is expressed as a function of density and strain tensor $\mu(n, u_{\alpha\beta})$
\begin{equation}
d\mu = \frac{\partial \mu}{\partial n}\Big|_{u_{\alpha\beta}}dn + \frac{\partial \mu}{\partial u_{\alpha\beta}}\Big|_n du_{\alpha\beta} 
\end{equation}
which yields
\begin{equation}
\frac{\partial n}{\partial u_{\alpha\beta}}\Big|_\mu = -\Big(\frac{\partial \mu}{\partial n}\Big|_{u_{\alpha\beta}}\Big)^{-1} \frac{\partial \mu}{\partial u_{\alpha\beta}}\Big|_{n} 
 = n_0 \nu^{-1} \mu_{\alpha\beta}.
\end{equation}
The last two terms are
\begin{align}
\frac{\partial h_{\gamma\delta}}{\partial u_{\alpha\beta}}\Big|_\mu &= \frac{\partial h_{\gamma\delta}}{\partial u_{\alpha\beta}}\Big|_n + \frac{\partial h_{\gamma\delta}}{\partial n}\Big|_{u_{\alpha\beta}}\frac{\partial n}{\partial u_{\alpha\beta}}\Big|_\mu
\\  &= C^n_{\alpha\beta\gamma\delta} - \mu_{\alpha\beta}\nu^{-1}\mu_{\gamma\delta},
\\
\frac{\partial h_{\gamma\delta}}{\partial \mu}\Big|_{u_{\alpha\beta}} &= \frac{\partial h_{\gamma\delta}}{\partial n}\Big|_{u_{\alpha\beta}}\Big( \frac{\partial \mu}{\partial n}\Big|_{u_{\alpha\beta}}\Big)^{-1} 
\\ &= -n_0 \nu^{-1} \mu_{\gamma\delta}.
\end{align}
Now the alternative formula \eqref{kappa_alt} can be expressed with $\nu,\mu_{\alpha\beta},$ and $C^n_{\alpha\beta\gamma\delta}$, which yields Eq.~\eqref{eq43}.\\

The same procedure can be applied to $\kappa^c$, which leads to
\begin{equation}
\kappa^c=-\frac{1}{n_0^2}\frac{\partial c}{\partial \mu}\Big|_{u_{\alpha\beta}}\!\!\!+\frac{1}{n_0^2}\frac{\partial c}{\partial u_{\alpha\beta}}\Big|_{\mu}\left(\frac{\partial \sigma_{\gamma\delta}}{\partial u_{\alpha\beta}}\Big|_\mu\right)^{-1}\frac{\partial \sigma_{\gamma\delta}}{\partial \mu}\Big|_{u_{\alpha\beta}} \;  ,
\end{equation}
which is an alternative to Eq.~\eqref{defectcomp2}. 
The following connection to the elastic constants
\begin{align}
-\frac{1}{n_0^2}\frac{\partial c}{\partial \mu}\Big|_{u_{\alpha\beta}}&=\nu^{-1},\\
\frac{\partial c}{\partial u_{\alpha\beta}}\Big|_\mu &= -\Big(\frac{\partial \mu}{\partial c}\Big|_{u_{\alpha\beta}}\Big)^{-1} \frac{\partial \mu}{\partial u_{\alpha\beta}}\Big|_{c}=-n_0 \nu^{-1} \mu^c_{\alpha\beta},\\
\frac{\partial \sigma_{\gamma\delta}}{\partial u_{\alpha\beta}}\Big|_\mu &=C^c_{\alpha\beta\gamma\delta} - \mu^c_{\alpha\beta}\nu^{-1}\mu^c_{\gamma\delta},\\
\frac{\partial \sigma_{\gamma\delta}}{\partial \mu}\Big|_{u_{\alpha\beta}}&=-n_0 \nu^{-1} \mu^c_{\alpha\beta} \; ,
\end{align}
can be used, to  reproduce Eq.~\eqref{kappac}.

\section{Elasticity}\label{appElast}

With the expressions of Sect.~\ref{Kap3B}, the Eq.~\eqref{PerKorrU} reads
\begin{align}
\langle \delta u_\alpha^\ast \delta u_\beta\rangle &= Vk_BT \Lambda_{\alpha\beta}^{-1}(\mathbf{q}).
\end{align}
This follows from Eq.~\eqref{CorrNUMatrix2} with $V_{\alpha}(\mathbf{q})=0$, or $\mu^c=\nu\delta_{\alpha\beta}+\mu_{\alpha\beta}=0$ in the low $q$-limit, for a  crystal with vanishing coupling between strain and defects. If we assume the crystal to be ideal, viz.~defect free, then additionally the defect density correlations $\langle\delta c\delta c\rangle$ and with it $\kappa^c$ should be zero. This implies that $\nu(\mathbf{q})^{-1}$ vanishes, as follows from Eq.~\eqref{CorrNUMatrix2}. The correlations of the coarse-grained density Eq.~(\ref{InverseAll}a) then become
\begin{align}
\langle \frac{\delta n^\ast \delta n}{Vk_BT n_0^2}\rangle &= q_\alpha \Lambda^{-1}_{\alpha\beta}(\mathbf{q}) q_\beta = q_\alpha (C_{\alpha\epsilon\beta\phi} q_\epsilon q_\phi)^{-1} q_\beta,
\end{align}
where we used the small q expansion of the constants of elasticity and took care of the proper symmetric combination as discussed in [\onlinecite{Walz10}]. The elastic constants $C_{\alpha\beta\gamma\delta}$ of (ideal) elasticity theory correspond to $C^c_{\alpha\beta\gamma\delta}$ in Eq.~\eqref{ElastKonstCc}. In this ideal crystal approximation, the compressibity becomes
\begin{align}
\kappa=\delta_{\alpha\beta}(C_{\alpha\beta\gamma\delta})^{-1}\delta_{\gamma\delta} = \sum_{i,j=1}^3(C_{ij})^{-1}
\end{align}
Note, that the limit $\kappa^c=0$ and Eq.~\eqref{eq60} also imply $\kappa^\sigma=0$, while $\kappa$ has a finite limit,  showing the difference arising from the different constant stress tensors.

The elastostatic theory is contained in the static limit of the hydrodynamic equations of motion, see Eqs. (87) in Ref.~[\onlinecite{Walz10}]. Without point defects the only non-vanishing equation is (87c), which then reads
\begin{align}
q_\beta C_{\alpha\beta\gamma\delta} q_\delta u_\gamma &= 0.
\end{align}
But this is just the Fourier-transformed equation of elastostatics \eqref{GrundglElastostatik}.

\end{document}